\documentclass[journal,twocolumn,10pt]{IEEEtran}

\IEEEoverridecommandlockouts                              
\usepackage{steinmetz}
\usepackage{algorithm}
\usepackage{algpascal}
\usepackage{algpseudocode}
\usepackage[pdftex]{graphicx}
\usepackage[cmex10]{amsmath}
\usepackage{url}
\usepackage{color}
\usepackage{cite}
\usepackage{bm}
\usepackage[font = footnotesize]{caption}
\usepackage{subcaption}
\usepackage{lipsum}
\usepackage{epstopdf}
\usepackage{fancyhdr}
\usepackage[percent]{overpic}
\usepackage{multirow}
\usepackage{multicol}
\usepackage{comment}
\usepackage{graphics}
\usepackage{graphicx}
\usepackage{tikz}
\usepackage{amssymb, amsmath, amsbsy}
\usepackage{upgreek}
\usepackage{cancel}
\usepackage{color}
\usepackage{mathdots}
\usepackage{mathrsfs}
\usepackage{stackrel}
\usepackage[thinlines]{easytable}
\usepackage{physics}
\usepackage{upquote}
\usepackage{mathtools}
\usepackage{siunitx}
\usepackage{boldline}

\usepackage{booktabs}
\usepackage{soul}
\usepackage{makecell}
\DeclarePairedDelimiter\ceil{\lceil}{\rceil}
\definecolor{mycolor}{RGB}{0, 130, 0}

\begin{document}
\title{Piecewise Digital Predistortion for \\ mmWave Active Antenna Arrays: \\ Algorithms and Measurements}

\author{Alberto~Brihuega,~\IEEEmembership{Student Member,~IEEE,}
        Mahmoud~Abdelaziz,~\IEEEmembership{Member,~IEEE,}
        Lauri~Anttila,~\IEEEmembership{Member,~IEEE,}
        Matias~Turunen,
        Markus All\'{e}n,~\IEEEmembership{Member,~IEEE,}
        Thomas~Eriksson,~\IEEEmembership{Member,~IEEE,}
        and Mikko~Valkama,~\IEEEmembership{Senior~Member,~IEEE}

\thanks{The research work leading to these results was funded in part by the Academy of Finland (under the projects \#304147, \#301820, and \#319994), and in part by the Tampere University Doctoral School.}
\thanks{A.~Brihuega, L.~Anttila, M.~Turunen, M.~All\'{e}n, and M.~Valkama are with the Department of Electrical Engineering, Tampere University, Tampere, Finland.} 
\thanks{M.~Abdelaziz is with Zewail City of Science and Technology, Egypt}
\thanks{T.~Eriksson is with the Department of Electrical Engineering, Chalmers University, Gothenburg, Sweden.}

\vspace{-0mm}

}
\maketitle

\begin{abstract}
In this paper, we describe a novel framework for digital predistortion (DPD) based linearization of strongly nonlinear millimeter-wave active antenna arrays. Specifically, we formulate a piecewise (PW) closed-loop (CL) DPD solution and low-complexity gradient-adaptive parameter learning algorithms, together with a region partitioning method, that can efficiently handle deep compression of the PA units. The impact of beamsteering on the DPD performance is studied, showing strong beam-dependence, thus necessitating frequent updating of the DPD. In order to facilitate fast adaptation, an inexpensive, non-iterative, pruning algorithm is introduced, which allows to significantly reduce the number of model coefficients. The proposed methods are validated with extensive over-the-air RF measurements on a 64-element active antenna array transmitter operating at 28 GHz carrier frequency and transmitting a 400 MHz 5G New Radio (NR) standard-compliant orthogonal frequency division multiplexing waveform. The obtained results demonstrate the excellent linearization capabilities of the proposed solution, conforming to the new 5G NR requirements for frequency range 2 (FR2) in terms of both inband waveform quality and out-of-band emissions. {The proposed PW-CL DPD is shown to outperform the state-of-the-art PW DPD based on the indirect learning architecture, as well as the classical single-polynomial based DPD solutions in terms of linearization performance and computational complexity by a clear margin.}
  
\end{abstract}
\section{Introduction}
\IEEEPARstart{P}{ower}-efficient operation of transmitters (TXs) is of fundamental importance in any modern wireless system, and is also one of the key design criteria for 5G New Radio (NR) base stations (BSs) \cite{intro_1}. In general, millimeter-wave (mmWave) devices have inherently lower power efficiency than devices operating at lower frequencies, due to higher parasitic losses. Thus, to achieve good efficiency at mmWaves, and to obtain reasonable network coverage, highly nonlinear power amplifiers (PAs) operating close to saturation are expected to be utilized, which in turn gives rise to high levels of nonlinear distortion \cite{GreenComm,PA_design}. 
Such nonlinearities induce inband and out-of-band (OOB) distortions that may limit the capacity and quality-of-service of the intended and neighboring channel users. Thus, their levels are strictly governed by standardization bodies through different figures of merit (FoM), such as error vector magnitude (EVM), adjacent channel leakage ratio (ACLR) and spurious emission limits \cite{3GPPTS38104}. These FoMs are well understood in the context of single or few antenna TXs, which have traditionally operated in the sub-6 GHz (or frequency range 1, FR1, in NR terminology) part of the spectrum, and are characterized at the antenna ports. However, for array TXs operating in the recently standardized frequency range 2 (FR2) bands, 3GPP has defined new OOB emission limits, in addition to new procedures for quantifying them \cite{3GPPTS38104}. Specifically, the OOB emission limit has been relaxed from the $45$ dBc ACLR limit, applicable at FR1, to $26-28$ dBc at the FR2/mmWave bands, and is to be characterized now by means of over-the-air (OTA) measurements.

In general, such relatively low ACLR values imply or allow for a very nonlinear operation point of the PAs. 
Traditional digital predistortion (DPD) solutions developed for FR1 generally aim at reducing the ACLR from initial values of some $30$~dBc down to $50$~dBc or so (see for example \cite{GreenComm} for an overview), such that the $45$ dBc target is comfortably met and reasonably good power efficiency is achieved. However, to obtain a similar power efficiency at mmWaves, the operation point must be clearly more nonlinear.
Consequently, DPD solutions tailored for FR2 array TXs must be able to operate under very nonlinear conditions, aiming at reducing the OOB emissions from around $20$ dBc initial ACLR to the $30-35$ dBc range. Traditional DPD solutions are typically not designed to operate in such strongly nonlinear conditions, and will generally not produce good linearization results, as will be demonstrated in the measurement section of this paper. 

Interactions between antennas constitute another important challenge in the linearization of array TXs. In order to obtain a small form factor, isolators between the PAs and the corresponding antennas are preferably avoided \cite{Swedes_review}, and the mutual coupling between neighboring antennas causes the PAs' output port impedances and, consequently, their nonlinear behaviour to change with the steering angle \cite{Swedes_review}. This calls for reduced-complexity parameter learning solutions that enable fast adaptation of the DPD, since beams can be adapted in 5G NR networks at millisecond level. 

\subsection{Nonlinear Distortion in Array TXs -- State-of-the-Art}
Current state-of-the-art array linearization solutions focus on developing efficient processing and learning architectures to simultaneously linearize multiple and mutually different PAs \cite{DPD_DigitalMIMO,DPD_MM_4,DPD_MM_5,DPD_MM_6,OTA_combining_DPD,Our_OTA_DPD}. These works consider a DPD learning signal that characterizes the combined signal at the receiver end. By doing so, a traditional single-input-single-output DPD learning problem is effectively obtained. Such DPD solutions result in minimizing the emissions in the main beam directions, where they have been shown to be most significant \cite{OOB_Mollen,OOB_Emissions_Ours}. Despite the works \cite{DPD_MM_4,DPD_DigitalMIMO,DPD_MM_5,DPD_MM_6,OTA_combining_DPD} constituting the basis of array linearization, they do not consider practical TXs, e.g., the potential crosstalk or load modulation effects are not accounted for, nor do they provide RF measurements with real array transmitters or strongly nonlinear operation points.

In order to effectively linearize the TX under crosstalk, \cite{MIMO_DPD_2,MIMO_DPD_3,MIMO_DPD_4} propose different multi-dimensional polynomial-based models for the task, since the performance of classical single-input models is clearly degraded in such scenarios. However, the complexity of the multi-dimensional models grows exponentially with the number of TX chains, and thus they are typically only tested with a small number of TX chains. To alleviate the complexity issue, \cite{Crosstalk_Chalmers} proposed instead a dual-input DPD architecture along with a proper crosstalk model that allows to linearize the TX efficiently with reasonable complexity.

The works \cite{Full_angleDPD,reduced_set_DPD,Our_OTA_DPD,OTA_DPD1,OTA_DPD2,OTA_DPD3} supported and confirmed through OTA measurements the basic theory provided in \cite{DPD_MM_4,DPD_MM_5,DPD_MM_6,OTA_combining_DPD} in the context of linearization of beamforming transmitters. In all these works, the reference signal for DPD learning was obtained through measurements from a far-field test receiver, and a conventional ILA-based DPD with least-squares (LS) estimation was considered. In \cite{Full_angleDPD}, the authors proposed to introduce tuning boxes that compensate for the potential mismatches between PAs, so that they all exhibit the very same behavior. By doing so, it is possible to provide linearization in all directions with a single DPD, as opposed to linearizing the main beam only. However, the tuning boxes are to be implemented in the analog domain, and they need to be estimated sequentially, introducing significant complexity and delay when large array TXs are considered. Furthermore, the potential changes in the PAs' behaviors due to crosstalk may have been overlooked due to the reduced array size as well as relatively low transmit power and bandwidth considered in \cite{Full_angleDPD}. Specifically, during the online operation, the main beam is pointed towards the intended receivers, thus the direct observation of the main beam signal from far-field test receivers is not necessarily possible. Hence, \cite{OTA_DPD2, OTA_DPD3} proposed different methods to reconstruct the main beam signal from sidelobe observations. However, potential beam-dependent load modulation was not considered in these works, limiting their real-life applicability.

In \cite{reduced_set_DPD,Our_OTA_DPD}, the experiments were conducted with a more compact and practical 64-element active array TX, and with the focus on linearizing the main-beam direction. In \cite{reduced_set_DPD} the implications that the load modulation has on the linearization performance were elaborated on, concluding that the DPD must be updated as the beam-direction changes. The study in \cite{Our_OTA_DPD} provides a measured proof of concept of a single DPD unit being capable of linearizing a beamforming array transmitter, and establishes the starting point for this work. The main limitation in \cite{Full_angleDPD,reduced_set_DPD,Our_OTA_DPD, OTA_DPD1,OTA_DPD2,OTA_DPD3} was the fact that the experiments were conducted such that the ACLR without DPD already fulfills the FR2 specifications, i.e., they considered a mildly nonlinear operation point. Furthermore, the ACLR numbers in \cite{Full_angleDPD,reduced_set_DPD,Our_OTA_DPD, OTA_DPD1,OTA_DPD2,OTA_DPD3} were not calculated through the total radiated power (TRP) as defined in \cite{3GPPTS38104} for transmitters operating at FR2, thus the reported numbers do not necessarily represent realistic standard-compliant ACLR performance.

As we will demonstrate in Section \ref{Measurements}, the performance of ILA-based techniques will largely degrade when the active array is operating close to compression. Additionally, the coefficient estimation in ILA is known to be prone to noise \cite{Morgan_noise} and to any linear distortion along the observation path, including limiting the observation bandwidth \cite{CL_vs_ILA,CL_limitedBW}. These effects are more severe in mmWave systems due to the wide bandwidths and the inherently higher insertion losses and noise figures of the devices. Closed-loop (CL) learning, on the other hand, is robust against these effects 
\cite{CL_vs_ILA}. Furthermore, as CL can tolerate noise better, a lower resolution ADC could in principle be utilized to further simplify the observation receiver. All these features make CL learning very appealing for DPD in mmWave systems, and thus we also consider and focus on CL learning in this article.

\subsection{Novelty and Contributions}
In this article, building partially on our initial early work in \cite{PW_Dec_DPD}, we describe efficient DPD processing and learning solutions for linearizing strongly nonlinear mmWave active arrays. The main contributions of the article can be described and summarized as follows:
\begin{itemize}
    \item {We formulate a  piecewise (PW) CL DPD structure along with an efficient parameter learning entity that are together able to successfully linearize a PA or an array of PAs under very nonlinear conditions. 
    The proposed solution is shown to outperform the early work in \cite{PW_Dec_DPD} and the widely adopted ILA-based single-polynomial DPD in \cite{Full_angleDPD,reduced_set_DPD,Our_OTA_DPD,OTA_DPD1,OTA_DPD2,OTA_DPD3} by a wide margin. } 
    \item {For proper PW modelling, we propose a novel region partitioning algorithm, specifically tailored for PA modeling and linearization. When adopted in the context of the devised PW-CL DPD system, enhanced linearization performance is shown, compared to the state-of-the-art PW-ILA DPD in \cite{PW_chalmers}.}
    \item To facilitate reduced complexity parameter learning and predistortion, a pruning algorithm that selects the most relevant basis functions (BFs) based on their contribution to the residual error is proposed. The algorithm leverages information that is inherently available in the proposed CL algorithm, and thus entails minimum additional complexity.
    \item All the measured ACLR evaluations build on the new 3GPP 5G NR specifications for FR2 \cite{3GPPTS38104}, being among the first in the open DPD literature. The experiments are performed through OTA measurements utilizing a 64-element active antenna array TX operating at 28 GHz carrier frequency and transmitting a 5G NR carrier with 400 MHz instantaneous bandwidth and NR standard-compliant orthogonal frequency division multiplexing (OFDM) waveform.
    \item The proposed solution is shown to be able to linearize the active antenna array under very strong nonlinear conditions, with initial ACLR levels as low as 21 dBc, whereas the current state-of-the-art techniques cannot provide anymore sufficient linearization. By using the proposed technique, the EIRP can be increased by more than 4 dB compared to the reference techniques, when considering 5G NR specifications for ACLR and EVM as a benchmark. This allows for greatly improved energy efficiency and network coverage.
\end{itemize}

We also elaborate on the implications of the load-modulation stemming from the antenna coupling on the DPD performance, and analyze the behavior of the OOB emissions in the spatial domain. 
Our experimental results with a state-of-the-art mmWave active array also support the notion that linearizing only the main-beam direction is sufficient in practical scenarios, though it is also noted that this is always subject to the specific array hardware. 
Finally, while the main performance evaluations are utilizing the mmWave active array, the applicability of the proposed closed-loop DPD system and region partitioning methods is also demonstrated in the context of linearizing a strongly nonlinear GaN {Doherty} PA operating at 1.845~GHz  center frequency (NR band n3). 

The rest of the paper is organized as follows. Section \ref{sec:nonlinear_model} provides the nonlinear modelling of practical beamforming mmWave TXs. In Section \ref{sec:proposed_DPD}, the proposed PW-CL DPD structure along with the parameter learning algorithm are introduced. {The proposed region partition algorithm is introduced and described in Section \ref{sec:region_partition}, together with a corresponding RF measurement example, while the proposed pruning algorithm is described in Section \ref{sec:pruning_algorithm}.} Section \ref{sec:Complexity} provides an extensive complexity analysis of the proposed PW-CL DPD and its comparison against several reference solutions. The conducted mmWave OTA experiments are provided and analyzed in Section \ref{Measurements}. Lastly, Section \ref{sec:Conclusions} provides the main concluding remarks.

\begin{figure}[t!]
\centering
\includegraphics[width=1\linewidth]{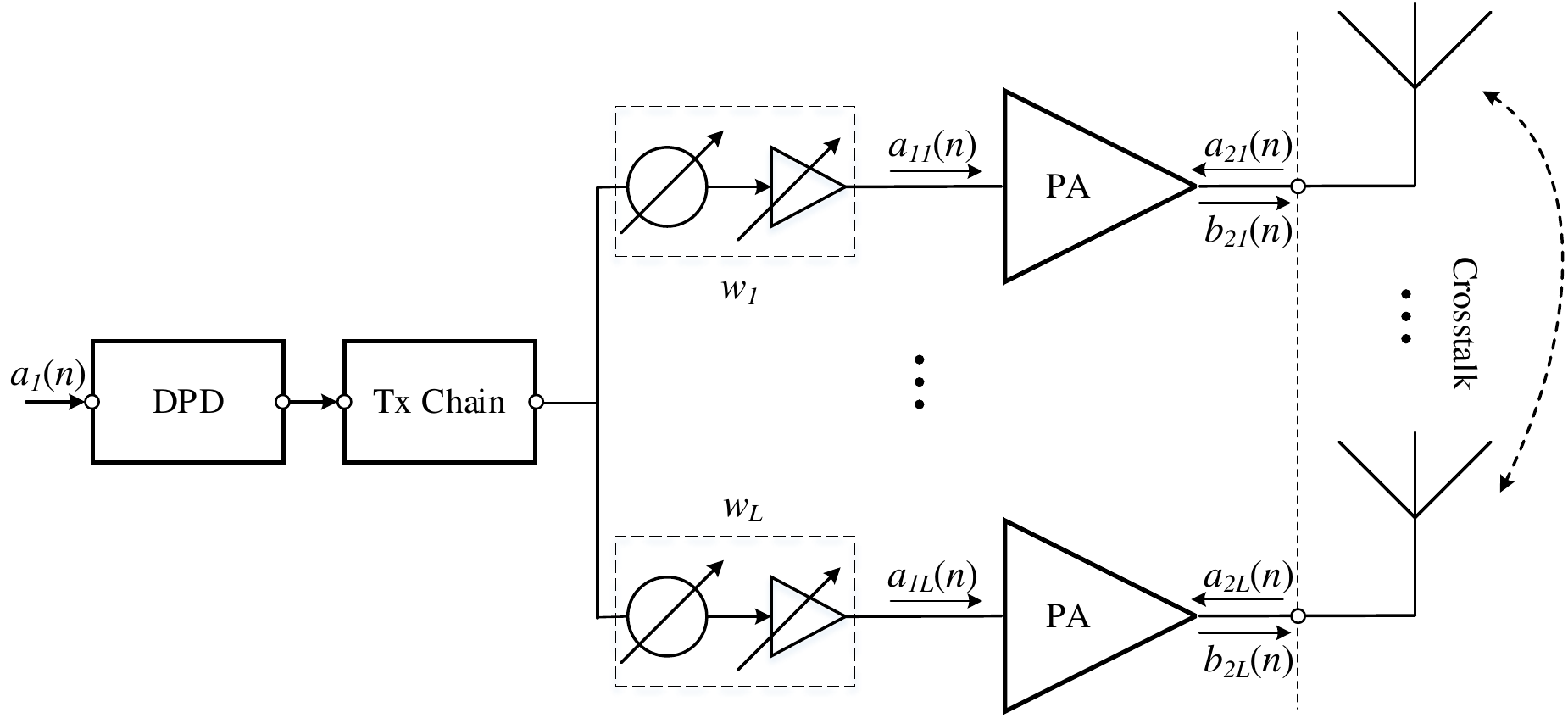}
\caption{\quad {Block diagram of the considered active array transmitter architecture.}}

\label{fig:Tx}
\end{figure}

\section{Nonlinear Array System Model}\label{sec:nonlinear_model}
In this section, the nonlinear model of the considered active array transmitter architecture depicted in Fig.~\ref{fig:Tx} is formulated. For notational convenience, we express the model in the equivalent discrete-time low-pass domain. {We also note that in this article, we focus on array transmitter systems that contain only a single transmit chain, as shown in Fig.~\ref{fig:Tx}.}  
\subsection{Transmitter Model}
With reference to Fig.~\ref{fig:Tx}, assume first that the DPD unit is off, and let $a_{1i}(n)=w_ia_1(n)$ denote the incident wave to the input of the $i$th PA, where $a_1(n)$ is the transmit I/Q waveform and $w_i$ is the analog beamforming coefficient, which may also account for windowing or tapering to control the sidelobes. Due to electromagnetic coupling between the antenna elements, the waves fed to the antennas also drive the output ports of the PAs, which is graphically illustrated in Fig. \ref{fig:Tx}. This results in an apparent dynamic variation of the PAs' output load, commonly referred to as load modulation, that depends on the steering angle \cite{Swedes_review}. In order to accurately model the behavior of the PAs under such conditions, we consider a dual-input behavioral model \cite{Prediction_distortion,Dual_input_model}, where the output signals $b_{2i}(n)$ are nonlinear functions of the incident waves $a_{1i}(n)$ and $a_{2i}(n)$. {Since only a single transmit stream or signal $a_1(n)$ is considered, it is possible to express the dual input behavioral model in \cite{Prediction_distortion,Dual_input_model} as a nonlinear function of $a_1(n)$ only \cite{reduced_set_DPD}. This essentially yields}

\begin{equation}
    \begin{split}
        b_{2i}(n) &= \sum_{\substack{m_1=0}}^{M_1}\sum_{\substack{p=0}}^{(P_1-1)/2}\alpha^{(2p+1)}_{m_1}w_i|w_i|^{2p}a_{1}(n-m_1)\\
        &\times|a_{1}(n-m_1)|^{2p}\\
        &+\sum_{\substack{m_2=0}}^{M_2}\beta^0_{m_2}f_i(n-m_2)\star a_1(n-m_2)\\
        &+ \sum_{\substack{m_3=0}}^{M_3}\sum_{\substack{m_4=0}}^{M_4}\sum_{\substack{p=1}}^{(P_2-1)/2}\beta^{2p+1}_{m_4m_3}|w_i|^{2p}\\
        &\times f_i(n-m_3)\star a_1(n-m_3)|a_{1}(n-m_4)|^{2p}\\
        &+\sum_{\substack{m_5=0}}^{M_5}\sum_{\substack{m_6=0}}^{M_6}\sum_{\substack{p=1}}^{(P_3-1)/2}\zeta^{2p+1}_{m_6m_5}w_i^2|w_i|^{p-1}\\
        &\times f^*_i(n-m_5)\star a^*_1(n-m_5)(a_{1}(n-m_6))^2\\
        &\times |a_{1}(n-m_6)|^{2(p-1)}\label{eq:dual_input_model_final},
    \end{split}
\end{equation}
{where $P_1, P_2, P_3$} denote polynomial orders and {$M_1, \hdots, M_6$} designate memory depths of the model, while $\alpha^{(2p+1)}_{m_1}$, $\beta^0_{m_2}$, $\beta^{2p+1}_{m_4m_3}$ and $\zeta^{2p+1}_{m_6m_5}$ are the model coefficients or free parameters. {Additionally, $f_i(n) = \sum_{\substack{l=1}}^{L}w_l\lambda_{il}(n)\star\mu_l(n)$ where $\lambda_{il}(n)$ is the filter impulse response that models the crosstalk from the $l$th to the $i$th antenna, $\mu_l(n)$ is an impulse response that models the linear distortion in the $l$th antenna/PA branch, $\star$ denotes the convolution operator, while $L$ is the total number of antennas.} 
{It is noted that even though the model is fairly complicated and nonlinear in the input samples, it is linear in the model parameters $\alpha^{(2p+1)}_{m_1}$, $\beta^0_{m_2}$, $\beta^{2p+1}_{m_4m_3}$ and $\zeta^{2p+1}_{m_6m_5}$, or alternatively in the effective model parameters where also the effects of the beamforming weights are lumped.}

In general, the signal model in (\ref{eq:dual_input_model_final}) has a relatively large number of BFs and coefficients, especially for larger values of the orders {$P_1, P_2, P_3$} and the memory depths {$M_1, \hdots, M_6$}. However, the model allows for any classical pruning alternative, e.g., reduction to a memoryless polynomial, memory polynomial (MP) or generalized memory polynomial (GMP) models, as discussed in \cite{reduced_set_DPD}. {In this article, instead of aiming at blindly pruning (\ref{eq:dual_input_model_final}), the information provided by the PW-CL DPD engine, which will be presented in Section \ref{sec:proposed_DPD}, is exploited in order to select the subset of the basis functions in (\ref{eq:dual_input_model_final}) that allows to meet a target residual distortion 
with the minimum amount of model coefficients. This is particularly applicable with orthogonalized basis functions, as described in further details in Section~\ref{sec:pruning_algorithm}.}

\subsection{Observation Model}

As the new FR2 signal quality metrics and the proposed DPD learning methods build on the combined OTA signal, either explicitly or implicitly through observation hardware, we next proceed with expressing the combined or observable signal and its effective nonlinear distortion. We consider LOS dominated propagation conditions and assume that the analog beamforming coefficients are chosen such that most of the energy is radiated towards the intended user's direction, i.e., $\phase{w_i} = \phase{h}^*_i$, where $h_i$  is the LOS component of the channel between the $i$th antenna and the intended user, while $\phase{x}$ refers to the phase angle of the argument variable $x$. The OTA combined signal at the intended user, in the absence of noise, can then be expressed as

\begin{equation}
    \begin{split}
    r(n) &= \sum_{\substack{i=1}}^{L}h_ib_{2i}(n)\\
     &= \sum_{\substack{m_1=0}}^{M_1}\sum_{\substack{p=0}}^{(P_1-1)/2}\bar{\alpha}^{(2p+1)}_{m_1}a_{1}(n-m_1)|a_{1}(n-m_1)|^{2p}\\
        &+\sum_{\substack{m_2=0}}^{M_2}\bar{\beta}^0_{m_2}(n-m_2)\star a_1(n-m_2)\\
        &+ \sum_{\substack{m_3=0}}^{M_3}\sum_{\substack{m_4=0}}^{M_4}\sum_{\substack{p=1}}^{(P_2-1)/2}\bar{\beta}^{2p+1}_{m_4m_3}(n-m_3)\\
        &\star a_1(n-m_3)|a_{1}(n-m_4)|^{2p}\\
        &+\sum_{\substack{m_5=0}}^{M_5}\sum_{\substack{m_6=0}}^{M_6}\sum_{\substack{p=1}}^{(P_3-1)/2}\bar{\zeta}^{2p+1}_{m_6m_5}(n-m_5)\\
        &\star a^*_1(n-m_5)(a_{1}(n-m_6))^2\times |a_{1}(n-m_6)|^{2(p-1)}\label{eq:rx_signal},
    \end{split}
\end{equation}
where $\bar{\zeta}^{2p+1}_{m_6m_5}(n-m_5)={\zeta}^{2p+1}_{m_6m_5}\sum_{\substack{i=1}}^{L}w_i^2|w_i|^{p-1}f^*_i(n-m_5)$, $\bar{\beta}^{2p+1}_{m_4m_3}(n-m_3) = \beta^{2p+1}_{m_4m_3}\sum_{\substack{i=1}}^{L}h_i|w_i|^{2p}f_i(n-m_3)$, $\bar{\alpha}^{(2p+1)}_{m_1} = \alpha^{(2p+1)}_{m_1}\sum_{\substack{i=1}}^{L}|w_i|^{2p+1}$ and $\bar{\beta}^0_{m_2}(n-m_2)=\beta^0_{m_2}\sum_{\substack{i=1}}^{L}h_if_i(n-m_2)$  are the effective model coefficients from the beamformed channel perspective. {We note that also (\ref{eq:rx_signal}) is linear in the effective model parameters, while the basis functions are the same as in (\ref{eq:dual_input_model_final}).}

From (\ref{eq:rx_signal}), it is possible to observe that the nonlinear behavior of the array depends on the actual coupling, which is here expressed through the filters $\lambda_{il}(n)$ built in $f_i(n)$. Such coupling is dependent on the beamforming direction, and will directly impact the observable nonlinear distortion. Its effect on the DPD performance is investigated with measurements in Section \ref{Measurements}.

\section{Proposed PW-DPD Structure and Parameter Learning Solution}\label{sec:proposed_DPD}

\begin{figure*}[t!]
	\centering
	\centerline{\includegraphics[width=0.82\textwidth]{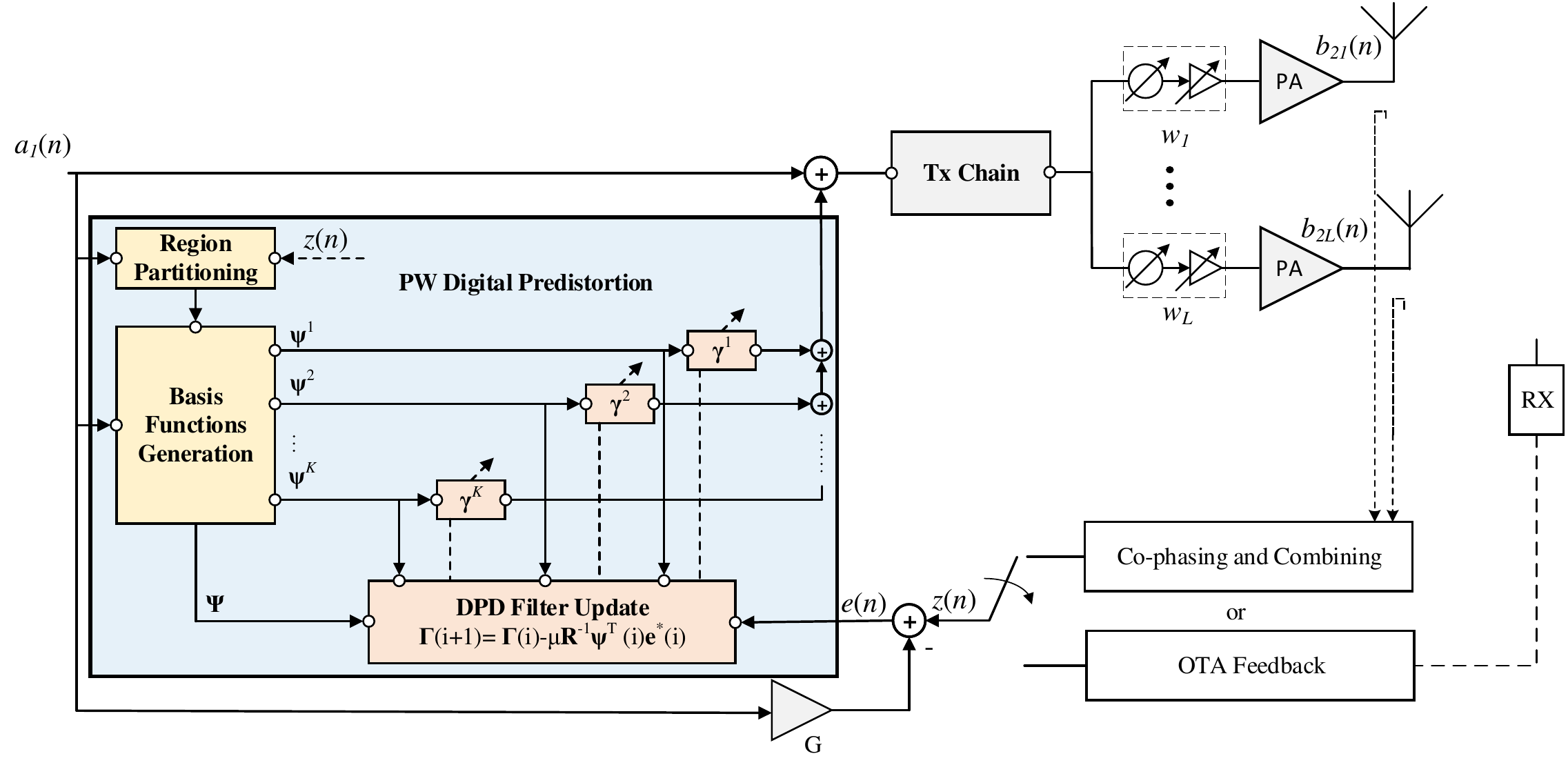}}
	\caption[]{\quad {Block diagram of the proposed closed-loop piecewise DPD solution with self-orthogonalizing gradient-adaptive parameter learning.}}
	\label{fig:PW_Decorr_DPD}
\end{figure*}

Global polynomial approximations are known to be sub-optimal when applied to the modeling of saturated or otherwise badly behaving functions. This global dependence on local effects can be largely avoided by using PW polynomial models \cite{deBoor_Splines}.
Now, since the new 5G NR emission limits at FR2 allow for very nonlinear operation, we propose and describe in the following a {new PW-CL DPD} solution that operates more robustly than global single-polynomial based models under such scenarios. The block diagram of the overall DPD solution is illustrated in Fig.~\ref{fig:PW_Decorr_DPD}. 

The parameter learning builds on mimicking or generating a local replica of the combined observable signal in (\ref{eq:rx_signal}), which captures the nonlinear distortion stemming from the active array from the beamformed channel perspective. It is noted that the proposed solution does not depend on the actual method of obtaining the learning signal, while a common approach is phase-aligning and combining the PA output signals \cite{DPD_MM_5,DPD_MM_6}. In a TDD system, the receiver-side beamformer could potentially be repurposed for this task. An alternative method is to send OTA feedback information or measurements from a remote test receiver \cite{Swedes_review, OTA_DPD1,OTA_DPD2,OTA_DPD3}. Both of these approaches are acknowledged, as alternatives, in  Fig. \ref{fig:PW_Decorr_DPD}.

\subsection{DPD Structure}

The basic idea is to inject into the digital transmit waveform a low power nonlinear signal with similar structure to the observable nonlinear distortion in (\ref{eq:rx_signal}), but with opposite phase, such that the nonlinear distortion cancels out at the receiver end \cite{DPD_MM_5}. Additionally, the nonlinear behavior of the TX is modeled in a piecewise manner, and therefore a different nonlinear function, referred to as a submodel \cite{PW_chalmers}, models an individual sub-region of the overall nonlinear response. 

Noting that the models in (\ref{eq:dual_input_model_final}) and (\ref{eq:rx_signal}) are linear-in-parameters, we denote the whole set of the global basis functions in these models as $\psi_j, j\!=\!1,2, \dots, B$, where $B$ is the total number of basis functions in the set. Then, considering the injection principle, the output signal of the piecewise DPD can be written as

\begin{equation}
    \Tilde{a}_1(n) =  a_1(n) + \sum_{\substack{k=1}}^{K}\sum_{\substack{j=1}}^{B_k} \gamma^k_{j} \psi^k_{j}(n) \label{DPD_out},
\end{equation}
where $K$ stands for the total number of submodels or regions of the PW model, $B_k$ is the number of basis functions in region $k$, $\gamma^k_{j}$ are the DPD coefficients, and  $\psi^k_{j}(n)$ are the PW basis functions, defined as 
\begin{align}
    \psi^k_{j}(n) = \begin{cases}
    \psi_j(n), & \text{if $u_k \leq \abs{a_1(n)} < v_k$} \\
    0, & \text{otherwise.}
    \end{cases}
\end{align}
Here, $u_k$ and $v_k$ denote the lower and upper limits of the $k$th region, respectively. {The exact way of calculating the regions, i.e., calculating $u_k$ and $v_k, k\!=\!1,2, \dots, K$, is described in detail in Section \ref{sec:region_partition}. It is also noted that the total number of coefficients per region can, in general, vary along the regions.}

\subsection{Gradient-based PW-Decorrelation DPD Learning}
The parameter learning is formulated such that the correlation between the observable nonlinear distortion in the main beam direction and the basis functions $\psi^k_{j}(n)$ is minimized. In the proposed solution, the adaptive learning is done jointly for all the regions, with the overall block diagram being illustrated in Fig. \ref{fig:PW_Decorr_DPD}. 

To this end, the combined learning signal at the output of the observation receiver is given by
\begin{align}
    z(n) &= \sum_{\substack{i=1}}^{L}e^{j\phase{w}^*_i} b_{2i}(n),
   \label{eq:observation_receiver}
\end{align}
which essentially yields (\ref{eq:rx_signal}) under pure LOS conditions. 
Alternatively, we can write $z(n)$ as
\begin{equation}
    {z(n) = G a_1(n) + d(n)},
\end{equation}
where {$G$} and $d(n)$ are the effective {complex linear gain} and the effective distortion term, containing linear and nonlinear distortion, respectively. The error signal that is used to adapt the DPD coefficients is then given by
\begin{equation}
    {e(n) = z(n) - \hat{G} a_1(n)}\label{eq:error_signal},
\end{equation}
where $\hat{G}$ refers to an estimate of $G$ and can be obtained, e.g., by means of least-squares fitting. 

The DPD coefficients are concatenated to a single vector as
\begin{align}
  \boldsymbol{\Gamma}(i) &= [\boldsymbol{\gamma}^1(i)\:\boldsymbol{\gamma}^2(i)\:\cdots\: \boldsymbol{\gamma}^K(i)]^T,
\end{align}
with
\begin{align}
 \boldsymbol{\gamma}^k(i) &= [\gamma^k_{1}(i)\:\cdots \gamma^k_{B_k}(i)],
\end{align}
and $i$ denoting the block index.
The data matrix, for an estimation block size of $N$ samples, is constructed as
\begin{align}
  \boldsymbol{\Psi}(i) = [\boldsymbol{\Psi}^1(i) \: \boldsymbol{\Psi}^2(i)\: \cdots\: \boldsymbol{\Psi}^k(i)\: \cdots\: \boldsymbol{\Psi}^K(i)],
\end{align}
Here, $\boldsymbol{\Psi}^k(i) \in \mathbb{C}^{N\times B_k}$ contain the PW basis functions for region $k$, and are defined as
\begin{align}
    \boldsymbol{\Psi}^k(i)=[\boldsymbol{\psi}^k_{1}(i)\: \cdots \: \boldsymbol{\psi}^k_{B_k}(i)],
\end{align}
with the basis function vectors
\begin{align}
    \boldsymbol{\psi}^k_{j}(i)=[\psi^k_{j}(n_i)\: \cdots \: \psi^k_{j}(n_i+N-1)]^T,
\end{align}
where $n_i$ denotes the starting sample index within the $i$th block.

In general, the BFs $\boldsymbol{\psi}^k_{j}$ are strongly mutually correlated, which can slow down the convergence of the DPD learning. Hence, orthogonalization of the basis functions can be adopted to ensure a fast and smooth convergence, especially with LMS-type algorithms. Such orthogonalization can be done with an orthogonal decomposition of $\boldsymbol{\Psi}$ through, e.g., Cholesky decomposition. As an alternative, to avoid explicitly transforming the BFs, a self-orthogonalized learning rule \cite{Haykin_AdaptiveFilters} can be adopted. The self-orthogonalized and classical block-adaptive learning rules are given, respectively, by 
\begin{equation}
    \boldsymbol{\Gamma}(i+1) = \boldsymbol{\Gamma}(i) - \mu\mathbf{R}^{-1}\mathbf{\Psi}^T(i)\mathbf{e}^*(i)\label{eq:learning_rule1}
\end{equation}
\begin{equation}
    \boldsymbol{\Gamma}_{\bot}(i+1) =  \boldsymbol{\Gamma}_{\bot}(i) - \mu\mathbf{\Psi}_{\bot}^T(i)\mathbf{e}^*(i), \label{eq:learning_rule2}
\end{equation}
where $\mu$ is the learning rate, 
$\mathbf{R} =\mathbb{E}\{ \boldsymbol{\psi}(n) \boldsymbol{\psi}(n)^H \}$ is the covariance matrix of the DPD input vector $\boldsymbol{\psi}(n)$ (a single row of $\boldsymbol{\Psi}(i)$), $\mathbf{e}(i) = [e(n_i)\: \cdots\: e(n_i+N-1)]^T$ represents the error vector containing the information of the prevailing nonlinear distortion in the main beam direction, and $\mathbf{\Psi}_{\bot}^T$ is the data matrix corresponding to orthogonalized BFs. The updated filter coefficients $\boldsymbol{\Gamma}(i+1)$ are then utilized to filter the next block of samples, and the processing is iterated until convergence. {A similar gradient-based learning rule is adopted in \cite{Decorr_learning_others}.}

{It is important to note that PW polynomials do not, in general, impose any continuity constraints between regions. However, since in the proposed PW-DPD system the learning for all the regions is done jointly, the learning algorithm inherently ensures continuity when minimizing the nonlinear distortion,  provided that it fully converges to the desired solution, as discontinuities are generally one source of nonlinear distortion. This will be illustrated with measurement examples in Section \ref{sec:region_partition}, and is one clear difference compared to the solution described in \cite{PW_Dec_DPD}}.

\section{Proposed Region Partition Algorithm}\label{sec:region_partition}

In \cite{PW_chalmers}, a Voronoi, or nearest neightbour, quantizer that minimizes the Euclidean distance between the centroids and data set was proposed for direct modeling and PA inversion. The authors considered the well-known $K$-means algorithm for finding the optimal partition over the PA input amplitude range. While this is a valid and simple approach, it exhibits some major limitations, which are listed below:
\begin{itemize}
    \item The actual nonlinear behavior of the PA is not considered. Thus, it is not possible to tell which is the optimal parameterization for each of the submodels, nor the optimal number of regions. This can lead to reduced modelling accuracy or unnecessarily high complexity.
    
    \item The $K$-means algorithm provides the optimal region partition in the Euclidean sense, that is, the region centroids are chosen such that their distance to the data set is minimized. By following this approach, the centroids tend to gather around the average amplitude value, where most of the samples are concentrated.  
    Due to the amplitude distribution of typical communication signals, samples around the average amplitude level, in general, excite the PA response fairly linearly, and thus can be accurately modeled with few or even one region. 
    Having closer centroids, i.e., narrower regions, in the high amplitude range would be more preferable, because there the nonlinear behavior is also more distinct.

\end{itemize}{}

\subsection{Proposed Algorithm}
Building on the above limitations, we propose a region partition algorithm that provides the optimal partition for a given submodel parameterization and desired modelling error, which are key parameters in PA modelling and linearization. 
The proposed algorithm is based on the Taylor's theorem, which allows to obtain an approximation to an arbitrary function in the vicinity of a point (the so-called region) with a certain accuracy depending on the order of the Taylor polynomial (submodel parameterization). 

The maximum approximation error of the Taylor's polynomial of order $Q_k$, over the real-valued function $f(a)$, is given by its Lagrange's remainder and reads \cite{Calculus}
\begin{equation}
    e_k = \frac{f_{k,\text{max}}^{(Q_k+1)}(u_k, v_k)}{(Q_k+1)!}(v_k-u_k)^{Q_k+1},
\end{equation}
where $f_{k,\text{max}}^{(Q_k+1)}(u_k, v_k)$ is the maximum of the $(Q_k+1)$th derivative of $f(a)$ in the interval from $u_k$ to $v_k$ (the region endpoints).
Therefore, the maximum width of the $k$th region, denoted as $\Delta_k = v_k-u_k$, given the modeling error $e_k$ and polynomial order $Q_k$, is given by
\begin{equation}
    \Delta_k = \sqrt[Q+1]{\frac{(Q_k+1)!e_k}{f_{k,\text{max}}^{(Q_k+1)}(u_k, v_k)}}. \label{eq:region_width}
\end{equation}

In the PA modeling or DPD context, the function $f(a)$ is the memoryless AM/AM response of the PA that can be accurately modeled with a high-order polynomial, which is also easily differentiable. Such memoryless polynomial-based function can be straight-forwardly estimated from the input signal $a_1(n)$ and observation signal $z(n)$, e.g., {through a LS fit}. The pseudocode of the region partition algorithm is given in Algorithm \ref{alg:Partition_algorithm}.

\begin{algorithm}[t!]
\caption{Region Partition Algorithm}\label{alg:Partition_algorithm}
\begin{algorithmic}[1]
\State \textbf{Input}: $a_{1}(n)$, $z(n)$, $e_k$ and $Q_k$ 
\State Set $j=1$, $k = 1$
\State Set  $u_1$ = 0, $v_1 = \text{max}|a_{1}(n)|$ and $\Delta_k(0)=0$
\State Extract $f(a_1)$ from $a_{1}(n)$, $z(n)$
\While{$u_k<\text{max}|a_{1}(n)|$}
\While{$\Delta_k(j)-\Delta_k(j-1)\leq \delta$}
\State Compute $f_{k,\text{max}}^{(Q_k+1)}(u_k, v_k)$ from $f(a_1)$
\State Compute $\Delta_k(j)$ according to (\ref{eq:region_width})
\State Set $v_k = u_k + \Delta_k(j)$
\State Set $j=j+1$
\EndWhile
\State Set $u_{k+1} = u_k + \Delta_k(j)$
\State Set $v_{k+1} = \text{max}|a_{1}(n)|$
\State Set $k = k+1$
\EndWhile
\State \textbf{return} $u_k$ and $v_k$ for $k =1, \cdots,K$
\end{algorithmic}
\end{algorithm}
The proposed algorithm results in wider regions when the amplitude response of the device under test is more linear, since it is accurately approximated with the chosen polynomial order over a wide range of input amplitudes. On the other hand, narrower regions are provided when nonlinearities are stronger. These properties are very important for PA modelling and predistortion, where a proper region partition along with its associated submodel parameterization can help to reduce the complexity of the predistortion task without compromising the linearization performance.
It is important to note that the region partition does only depend on the instantaneous envelope samples, while potential dependencies between regions, e.g., due to memory effects, are handled by the DPD filters. It is also noted that the region partition depends on the overall shape of the AM/AM response of the considered PA system, and thus one can expect that it is fairly stable over time. Thus, assuming that the region partition is originally calculated such that the PA is excited close to the maximum power, it is expected that the estimated partition can be utilized in the DPD system without systematically updating or recalculating it. However, for example larger changes in the center-frequency may also change the fundamental shape of the AM-AM response, and thus in such cases new region partition should be calculated. 

\subsection{Region Partition Demonstration with GaN Doherty PA}\label{GaN_Experiments}
In order to demonstrate the operation of the proposed region partition algorithm, we proceed to linearize a {GaN Doherty PA ({RFHIC} RTH18008S-30) operating at 1.845 GHz center frequency (NR band n3) and at  +39.5~dBm output power. The test signal is composed of three contiguously aggregated 5G NR component carriers, each with 20~MHz channel bandwidth, and utilizing 15~kHz subcarrier spacing and 64-QAM subcarrier modulation. Weighted-overlap-and-add (WOLA) \cite{IEEE_WOLA} processing is used to improve the bandlimitation of the digital transmit waveform. The complementary cumulative distribution function (CCDF) of the sample-level peak-to-average power ratio (PAPR) of the composite digital waveform, after iterative clipping and filtering, reads 6.4~dB and 7~dB when measured at $1\%$ and $0.01\%$ points, respectively}. A National Instruments PXIe-5840 vector signal transceiver, which includes both a vector signal generator (VSG) and a vector signal analyzer (VSA), is utilized in this experiment.  
The VSG output is connected to a driver amplifier, which feeds the GaN {Doherty} PA, whose output is then connected to the VSA through a high power attenuator. PAs based on GaN {Doherty} technology typically exhibit strong amplitude dependent nonlinear behaviour. This makes its modeling and inversion through traditional single-polynomial based approaches quite challenging, thus allowing to efficiently leverage the capabilities of PW polynomials. For DPD parameterization, we consider the DPD structure described in (\ref{DPD_out}), but utilize generalized memory polynomial \cite{GMP_morgan} basis functions in this example, with memory depth of $3$, {cross memory depth of $2$} and nonlinearity order $5$ for all submodels. The simpler model is utilized here simply to demonstrate the operation of the region partition algorithm, and to provide a fair comparison against the reference solutions.

\begin{figure}[t!]
    \centering
    \begin{subfigure}[t]{0.8\linewidth }
        \includegraphics[width=\linewidth]{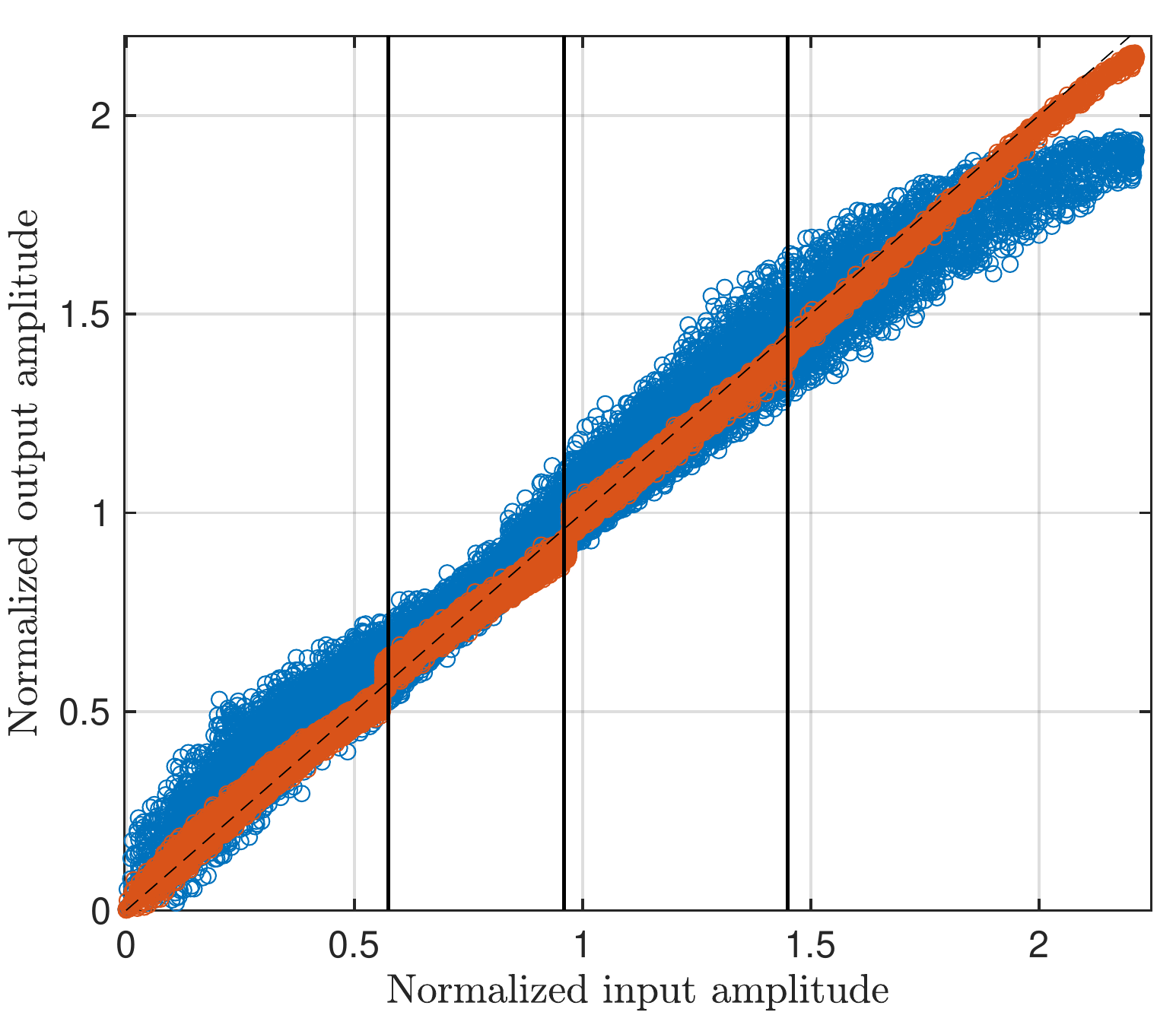}
    \end{subfigure}
    \begin{subfigure}[t]{0.8\linewidth}
        \includegraphics[width=\linewidth]{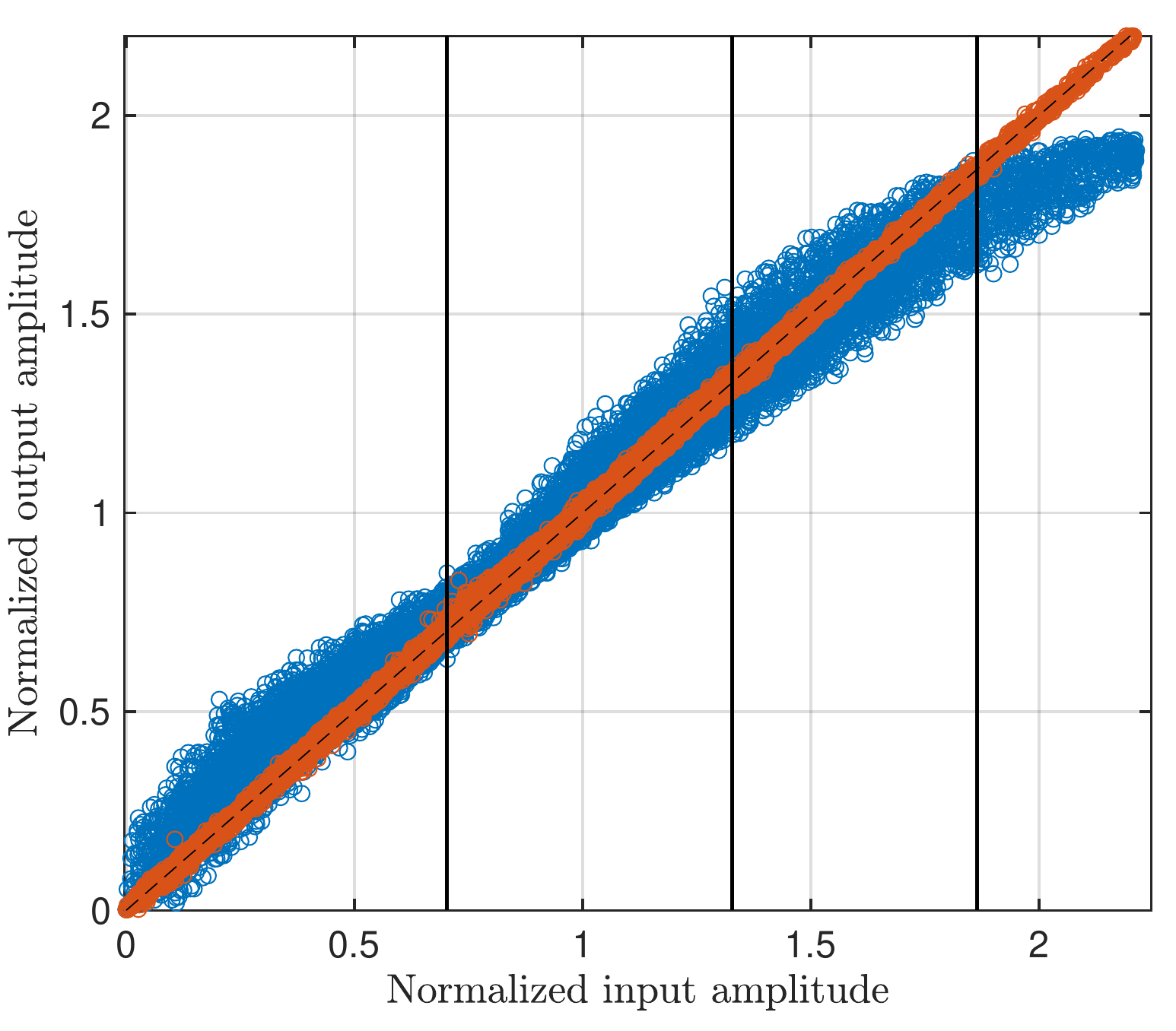}
    \end{subfigure}
    \vspace{0mm}
    \caption{\quad {AM/AM responses of GaN Doherty PA at 1.8~GHz band, measured with 3$\times$20~MHz NR waveform at an output power of $+39.5$~dBm. In top, the region partition provided by \cite{PW_chalmers} and the linearized response of PW-CL DPD in \cite{PW_Dec_DPD} are shown. In bottom, the region partition provided by Algorithm \ref{alg:Partition_algorithm} and the linearized response of the proposed PW-CL DPD are shown. Reference black dashed-line represents the ideal linear response.}} \label{fig:region_partition}
\end{figure}

The region partitions provided by the reference solution in \cite{PW_chalmers} and the proposed Algorithm \ref{alg:Partition_algorithm} are depicted in Fig. \ref{fig:region_partition}. To obtain the bottom partition, the submodel parameterization was set to $Q_k = 5$ with $e_k =0.01$ $\forall k$, resulting in $K=4$ regions. The number of regions of the reference solution was then also set to four. As it can be clearly observed, the reference solution yields a narrow region around the average amplitude of the signal, which could be over-parameterized, while the last region is the widest one and might benefit from a higher polynomial order for proper inversion. On the other hand, the proposed algorithm provides more evenly distributed regions so that a $5$th order polynomial can be easily utilized for inversion.

\begin{figure}[t!]
\centering
\includegraphics[width=1\linewidth]{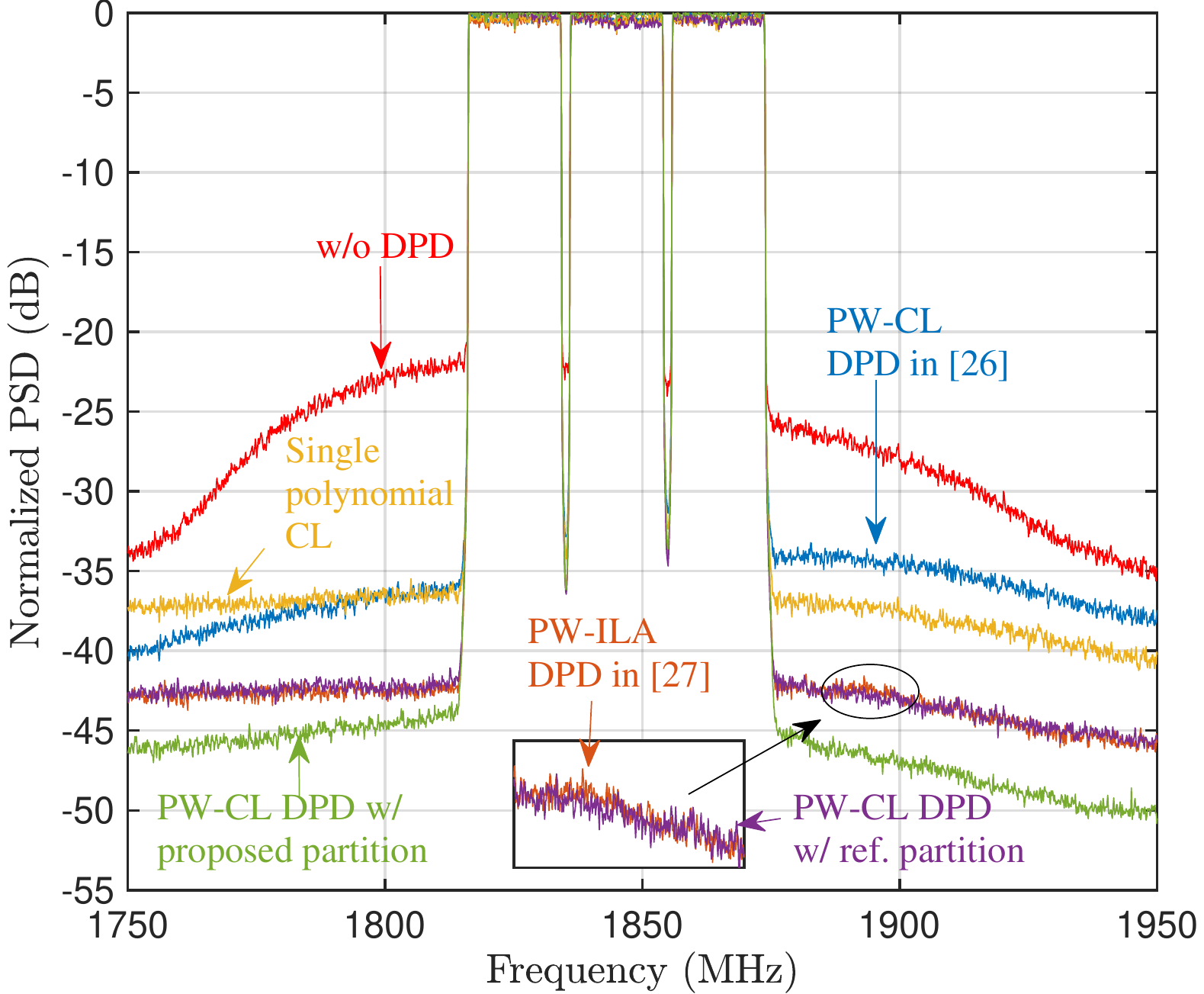}
\caption{{\quad Measured spectra at the GaN Doherty PA output, at NR band n3, when excited with three 5G NR component carriers each being 20~MHz wide. The PA output power is +39.5~dBm.}}
\label{fig:GaN_PA_spectra}
\end{figure}

In {Fig.~\ref{fig:region_partition}, the linearized AM/AM responses provided by the reference PW-CL DPD solution from \cite{PW_Dec_DPD} and by the new proposed PW-CL DPD are illustrated.  As it can be observed, the solution from \cite{PW_Dec_DPD} results in non-smooth transitions between the regions, and is thus incapable of linearizing the PA properly -- something that can be clearly observed also in Fig. \ref{fig:GaN_PA_spectra} showing the linearized spectra obtained through the different methods. The reduced linearization performance of the method in \cite{PW_Dec_DPD} is essentially stemming from the way how the piecewise DPD parameter learning and the handling of the associated BF correlation were done. Specifically, the orthogonalization transformation in \cite{PW_Dec_DPD} was done on the original non-piecewise BF matrix, and consequently, as the BF samples are split into $K$ regions, the basis functions within a region are no longer orthogonal. The remaining mutual correlation between the BFs greatly impacts the parameter learning with gradient/LMS like methods, preventing the learning system from converging properly, thus degrading the performance. 

On the other hand, as can be observed in Fig.~\ref{fig:region_partition}, the new proposed PW-CL DPD results into smooth transitions between regions. This is because the parameter learning for the piecewise DPD system is formulated differently, and the associated basis function orthogonalization or the use of self-orthogonalized learning are adopted such that the structure of the PW processing is properly embedded into the basis function matrix. Therefore, the learning seeking to minimize the distortion will implicitly result into continuity between the regions while converging to the steady-state solution. }

{The linearization performance, in terms of the PA output spectra, provided by the different DPD solutions is depicted in Fig.~\ref{fig:GaN_PA_spectra}. By examining Fig.~\ref{fig:GaN_PA_spectra}, it can be clearly observed that the proposed PW-CL DPD solution outperforms {the reference single polynomial CL, due to its limited capability of modelling such strong local nonlinear effects, as well as the reference PW-CL from \cite{PW_Dec_DPD} by a large margin}. The proposed PW-CL with the reference partition in \cite{PW_chalmers} provides the same linearization performance as the PW-ILA DPD, while utilizing the proposed region partition algorithm allows for a better modeling accuracy, in addition to providing the number of regions and the maximum nonlinearity orders $Q_k$, making thus the proposed region partition algorithm an efficient and appealing solution for PA inversion and modelling. Further experiments with the mmWave active array will be provided in Section~\ref{Measurements}.}

\section{Pruning Algorithm}\label{sec:pruning_algorithm}
{Large amount of basis functions and corresponding coefficients in the DPD system are known to be challenging from the processing complexity and parameter estimation points of view.}
In order to tackle this issue, various pruning algorithms have been proposed in the literature, in general. Many of these have been based on feature selection, e.g., LASSO \cite{LASSO} and ridge regression \cite{Ridge} that are based on $\ell_1$ and $\ell_2$ norms, respectively, or compressed sensing methods such as orthogonal matching pursuit (OMP) \cite{OMP_DPD} or doubly-OMP \cite{DOMP} that select in every iteration the regressor with the largest projection onto the residual. Another family of algorithms rely on feature extraction, considering only the input data to generate the new BF matrix of relevant components through, e.g.,  principal component analysis \cite{PCA_DPD,PCA_DPD2}. These algorithms allow to effectively reduce the number of model parameters without compromising the performance. However, the latency and complexity involved in their iterative operation may prevent fast adaptation. Hence, we seek computationally more efficient methods.

\subsection{Proposed Pruning Algorithm}
In this work, we propose a pruning methodology that resembles  \cite{Adaptive_BFs}, however, with some important distinctions. In \cite{Adaptive_BFs}, the algorithm selects the most relevant BFs during the actual DPD learning. The BFs are selected from an initial large set of BFs, which will be denoted here as $\boldsymbol{\Psi}^0$. Then, the contribution of each of the BFs towards improving the DPD linearization in the $i$th iteration of the learning is measured with $\abs{\boldsymbol{\zeta}(i)} = \abs{\boldsymbol{\Psi}^{0}(i)^T\mathbf{e}^*(i)}$, where the absolute value is taken element-wise. This quantifies the magnitude of the cross-correlations between the residual error signal and the basis functions. Then, two sets of BFs are defined, the so-called active set ${\boldsymbol{\Psi}}_{\text{ACT}}$, and the predistortion set $\boldsymbol{\Psi}_{\text{PD}}$. To this end, ${\boldsymbol{\Psi}}_{\text{ACT}}$ is composed of the BFs that fulfill a dynamic threshold $\boldsymbol{\zeta}(i)>\zeta_{\text{TH}}(i)$, and their associated filter coefficients are the only ones being updated. On the other hand, the predistortion set $\boldsymbol{\Psi}_{\text{PD}}$ is made up of the BFs that belong or have belonged to ${\boldsymbol{\Psi}}_{\text{ACT}}$ in previous iterations, and are the ones used to predistort the signal. The threshold $\zeta_{\text{TH}}(i)$, that controls which BFs are considered for learning and predistortion, becomes increasingly smaller in every iteration. As a result, new BFs are added to $\boldsymbol{\Psi}_{\text{ACT}}$ in each iteration, and consequently also to $\boldsymbol{\Psi}_{\text{PD}}$, and they need to be learnt from the scratch. This slows down the convergence speed of the DPD learning significantly, as reported also in \cite{Adaptive_BFs}, being thus potentially inadequate if fast adaptation is required. Furthermore, acquiring the pruning information, i.e., calculating $\boldsymbol{\zeta}(i)$ in each iteration requires additional computations.

\begin{algorithm}[t!]
\caption{Pruning Algorithm}\label{alg:Pruning_algorithm}
\begin{algorithmic}[1]
\State Define $\boldsymbol{\Psi}^0$ with BFs according to  (\ref{eq:dual_input_model_final})
\State {Orthogonalize $\boldsymbol{\Psi}^0$ to get $\boldsymbol{\Psi}^0_{\bot}$}
\State Set $\zeta_{\text{TH}}$ 
\While{\text{learning DPD}}
\If{i=1}
   \State $\boldsymbol{\Psi}_{\text{PD}}\gets \boldsymbol{\Psi}^0_{\bot}: \boldsymbol{\zeta}(1)>\zeta_{\text{TH}}$
\EndIf
\State  $\boldsymbol{\Psi}_{\text{ACT}}(i)\gets \boldsymbol{\Psi}_{\text{PD}}:\boldsymbol{\zeta}_{\text{ACT}}(i)>\zeta_{\text{TH}}$
\State  $ \boldsymbol{\Gamma}_{\text{ACT}}(i+1) =  \boldsymbol{\Gamma}_{\text{ACT}}(i) - \mu\mathbf{\Psi}_{\text{ACT}}^T(i)\mathbf{e}^*(i)$
\State update $\boldsymbol{\Gamma}_{\text{PD}}(i+1) \in \boldsymbol{\Gamma}_{\text{ACT}}(i+1)$ according to step 8
\State predistort the next block with $\boldsymbol{\Gamma}_{\text{PD}}(i+1)$ {and} $\boldsymbol{\Psi}_{\text{PD}}$
\EndWhile
\end{algorithmic}
\end{algorithm}

Compared to \cite{Adaptive_BFs}, the proposed CL DPD engine has the required pruning information, $\boldsymbol{\zeta}(i)=\mathbf{\Psi}^T(i)\mathbf{e}^*(i)$, already available in every block iteration, as shown in (\ref{eq:learning_rule1}) and (\ref{eq:learning_rule2}). Thus, no additional computational cost is involved in acquiring it. Additionally, instead of a dynamic threshold, we determine a constant threshold $\zeta_{\text{TH}}$ that is used to select all the relevant BFs based on their contribution to the residual distortion $\mathbf{e}(i)$. Consequently, opposed to \cite{Adaptive_BFs} where the active set considers new BFs in every iteration, the proposed algorithm selects all the relevant BFs already in the first iteration of the learning, such that the convergence speed is not compromised. Then, as the crosscorrelations $\boldsymbol{\zeta}(i)$  become increasingly smaller during the DPD iterations, the BFs that fall below the threshold will be dropped from the active set, and their associated coefficients will no longer be updated, providing thus further complexity reduction. The proposed algorithm is formally defined and summarized in Algorithm \ref{alg:Pruning_algorithm}. 

\begin{table*}[]
\setlength{\tabcolsep}{2.57pt}
\renewcommand{\arraystretch}{1.4}
\caption{\textsc{{DPD main path processing complexity per sample and total DPD learning complexity}}}
\label{tab:complexity}
\begin{tabular}{ccccccc}
\hline
                  &  & \multirow{2}{*}{\textbf{PW-ILA DPD}} & \multicolumn{2}{c}{\textbf{CL DPD}} & \multicolumn{2}{c}{\textbf{Proposed PW-CL DPD}} \\
                  &  &                   &     Self orth.      &    Orth. BFs      &      Self orth.     &     Orth. BFs    \\  \clineB{1-7}{2.5}\\ & & & & & & \vspace{-6mm}\\
\multirow{3}{*}{\vspace{-6.0mm}\makecell{\textbf{DPD}}} & BF gen. &    $2\ceil*{\frac{N_{\text{IPW}}}{K}}-1$               &     ${2N_{\text{ISP}}-1}$      &    ${2N_{\text{ISP}}-1}$      &     $2\ceil*{\frac{N_{\text{IPW}}}{K}}-1$      &    $2\ceil*{\frac{N^{\text{Pr}}_{\text{IPW}}}{K}}-1$      \\ 
                  & Filt. & $8\ceil*{\frac{N_{\text{PW}}}{K}}-2$ & ${8N_{\text{SP}}-2}$ & ${8N_{\text{SP}}-2}$ & $8\ceil*{\frac{N_{\text{PW}}}{K}}-2$ & $8\ceil*{\frac{N^{\text{Pr}}_{\text{PW}}}{K}}-2$         \\
                  & Orth. & $-$ & $-$ & $2N^2_{\text{SP}}$ & $-$ & $2\ceil*{(\frac{N^{\text{Pr}}_{\text{PW}}}{K})^2}$    \\ & & & & & \vspace{-3mm} \\ \hline \vspace{-3mm} & & & & & \\
\multirow{2}{*}{\vspace{-8.2mm}\makecell{\textbf{Learning}}} & BF gen. & $I_{\text{ILA}}B_{\text{ILA}}(2\ceil*{\frac{N_{\text{IPW}}}{K}}+1)$ & $-$ & $-$ & $-$ & $-$          \\
                  & DPD est. & \makecell{$4I_\text{ILA}K\ceil*{\frac{N_{\text{PW}}}{K}}^2\times$ \\$(\ceil*{\frac{B_{\text{ILA}}}{K}}+\ceil*{\frac{N_{\text{PW}}}{3K}})$} & ${I_{\text{CL}}B_{\text{CL}}N_{\text{SP}}(4N_{\text{SP}}+6)}$ & ${I_{\text{CL}}N_{\text{SP}}(8B_{\text{CL}}+2)}$ & ${I_{\text{CL}}B_{\text{CL}}\ceil*{\frac{N_{\text{PW}}}{K}}(4\frac{N_{\text{PW}}}{K}+6)}$& \makecell{$8B_{\text{CL}}\ceil*{\frac{N_{\text{PW}}}{K}}+ 2\ceil*{\frac{N^{\text{Pr}}_{\text{PW}}}{K}}I_{\text{CL}}$\\$+8\ceil*{\frac{N^{\text{Pr}}_{\text{PW}}}{K}}B_{\text{CL}}(I_{\text{CL}}-1)$}      \\ \hline
\end{tabular}
\end{table*}

{For clarity, it is noted that this approach is specifically tailored for systems that demand real-time adaptation and pruning, by avoiding the iterative manner in which pruning algorithms generally operate. A proper selection of all the relevant BFs can be achieved in a single iteration by considering an orthogonal set of BFs, thus this pruning algorithm is mostly suitable for the learning rule described in (\ref{eq:learning_rule2}). On the other hand, if real-time adaptation is not required, any of the pruning methodologies described in the introduction of this section can basically be applied to (\ref{eq:learning_rule1}).}

\subsection{Determining the Threshold $\zeta_{\text{TH}}$}
In order to determine the threshold value, we briefly show that there is an intuitive connection between the power of the residual distortion $\mathbf{e}(i)$, the cross-correlations $|\boldsymbol{\zeta}(i)|$, and $\zeta_{\text{TH}}$. For simplicity, let us consider a 5th-order memoryless non-linear system, whose output under the excitation $a_1(n)$ reads
\begin{align}
    b_1(n)  & = \alpha_1\psi_1(n) + \alpha_3\psi_3(n) + \alpha_5\psi_5(n),
\end{align}
where the BFs are assumed to fulfill $\mathbb{E}\{\psi_i(n)\psi^*_j(n)\}=\delta_{i,j}(n)$, i.e., orthogonal BFs are considered.

Similar to (\ref{eq:error_signal}), but considering a perfect linear gain estimate for notational simplicity, i.e., {$\hat{G} = \alpha_1$}, the error signal reads
\begin{equation}
    e(n) = \alpha_3\psi_3(n) + \alpha_5\psi_5(n),
\end{equation}{}
whose average power is given by
\begin{equation}
\begin{split}
    D_e(n) &= \mathbb{E}\{e(n)e^*(n)\} \\
    &= |\alpha_3|^2+|\alpha_5|^2.
    \end{split}
\end{equation}

On the other hand, the absolute value of the cross-correlations between the BFs and the error signal reads
\begin{equation}
\begin{split}
    |\zeta_3| &= |\mathbb{E}\{\psi_{3}(n)e^*(n)\}| = |\alpha^*_3|
\end{split}
\end{equation}
\begin{equation}
\begin{split}
    |\zeta_5| &= |\mathbb{E}\{\psi_{5}(n)e^*(n)\}| = |\alpha^*_5|.
\end{split}
\end{equation}
Thus, the power of the error signal can now be expressed in terms of $\zeta_3$ and $\zeta_5$ as
\begin{align}
    D_e(n) &= |\zeta_3|^2 + |\zeta_5|^2.
\end{align}
Consequently, there is a simple connection between $\zeta_i$ and the total error power. The threshold $\zeta_{\text{TH}}$ will select the BFs in $\boldsymbol{\Psi}^0$ whose contribution to the nonlinear distortion is more than the given value, although it does not necessarily guarantee that the resulting total error power is below this limit.
Importantly, as it will be shown in Section \ref{Measurements}, the proposed algorithm allows to reduce the number of model coefficients significantly, without compromising the linearization performance, {and practically without extra computations, since the relevant complex correlation values are available through the LMS-like closed-loop learning algorithm directly.}

\section{Computational Complexity Analysis}\label{sec:Complexity}
{In this section, an extensive complexity analysis of the proposed PW-CL DPD solution employing the learning rules described in (\ref{eq:learning_rule1}) and (\ref{eq:learning_rule2}), and its comparison against the classical single-polynomial CL-DPD proposed in \cite{Concurrent_abdelaziz} and the PW ILA-based DPD proposed in \cite{PW_chalmers}, is conducted.} To this end, we consider the number of floating point operations (FLOPs) as the complexity metric. We also differentiate between DPD learning and actual DPD main path linearization processing. In general, the DPD main path is more critical from the complexity point of view, since it is to be run continuously and in real time along with the data transmission. On the other hand, the learning is to be executed when the characteristics of the antenna array or its operating point change. It is commonly argued that such characteristics remain unchanged over longer periods of time, however, if the load modulation due to beam-steering turns out to be critical, the nonlinear characteristics may change with each scheduling unit which can be only a few OFDM symbols. 
In such cases, fast real-time DPD adaptation is needed, which in turn makes the learning complexity as critical as that of the actual main path predistortion.

The obtained complexity expressions in terms of FLOPs are given in Table \ref{tab:complexity}, in generic form, where 
{the utilized notations are as follows. $\ceil*{\cdot}$ denotes the ceil operator, $N_{\text{ISP}}$ and $N_{\text{IPW}}$ stand for the total numbers of instantaneous BFs, i.e., the BFs provided by the model in (\ref{eq:dual_input_model_final}) when $m_1=m_2=\cdots =m_6 = 0$, for the single-polynomial and PW DPD structures, respectively. On the other hand, $N_{\text{SP}}$ and $N_{\text{PW}}$ refer to the cardinalities of the whole sets of BFs, i.e., those containing all the basis functions given by (\ref{eq:dual_input_model_final}), while $N^{\text{Pr}}_{\text{PW}}$ denotes the number of BFs after pruning. Furthermore, $B_{\text{CL}}$ and $B_{\text{ILA}}$ denote the estimation block-sizes of the CL and ILA learning, respectively, while $I_{\text{CL}}$ and $I_{\text{ILA}}$ denote the total numbers of block iterations of each DPD structure.} 
{Additionally, \textit{BF gen.} refers to the complexity stemming from generating the different BFs, while \textit{Filt.} refers to the complexity of applying the DPD filters to the BF samples, specifically, in the case of PW-CL, this corresponds to computing (\ref{DPD_out}). Finally, \textit{Orth.} refers to the complexity stemming from orthogonalizing $\mathbf{\Psi}$ to get $\mathbf{\Psi}_{\bot}$, while \textit{DPD est.} is the complexity stemming from estimating the DPD filter coefficients by iterating the learning rules in (\ref{eq:learning_rule1}), (\ref{eq:learning_rule2}), or by the applying LS estimation in the case of PW-ILA.}

For clarity, the more detailed assumptions adopted in the complexity analysis are summarized as follows:
\begin{itemize}
    \item {One complex multiplication is assumed to cost 6 FLOPs, while one complex-real multiplication and one complex sum both cost 2 FLOPs\cite{Flops}}. 
    \item {The complexity of the CL algorithms is evaluated for both the self-orthogonalized learning rule employing non-orthogonal BFs and the classical LMS-like learning rule employing orthogonal BFs, as described in (\ref{eq:learning_rule1}) and (\ref{eq:learning_rule2}), respectively. The former is indicated in the table as \textit{Self-orth.} while the latter as \textit{Orth. BFs}}. 
    \item {The CL DPD methods utilize the same BFs in both the DPD main path and in the parameter learning. Consequently, the learning can re-utilize the BFs already available from the main path transmission whenever needed. This is, however, not the case for the ILA architecture.}
     \item {The CL algorithms employing orthogonal BFs require the orthogonalization of the basis function matrix $\mathbf{\Psi}$. This is done through the Cholesky decomposition \cite{Matrix_computations}, utilizing the inverse matrix of the real-valued lower triangular matrix given by the decomposition. This inverse matrix is assumed to be precomputed, as it depends on the statistics of the transmit signal. Similarly, it is also assumed that the inverse of the correlation matrix in (\ref{eq:learning_rule1}) is precomputed.}
     \item {The PW-CL DPD method utilizing orthogonal BFs is assumed to use the pruned BFs. To this end, for the fairness of the complexity analysis, the learning algorithm in the first iteration has to calculate the product $(\mathbf{\Psi}_\bot^0)^T\mathbf{e}^*(n)$, while in the rest of $(I_{\text{CL}}-1)$ iterations, only the pruned BFs are processed. 
     }
     \item {For notational simplicity, it is assumed that all regions in the PW models have the same number of BFs. Then, as some of the involved operations work on a per region basis, the total number of PW BFs are divided by $K$ when applicable. For example, in the BF generation or filtering, a specific sample does only belong to a single region, and thus as a result, the generation and filtering only involves the operations associated to the BFs of such specific region.}
    \item {Generating the memory BFs does not involve any FLOPs, as they are delayed versions of the instantaneous BFs.}
    \item The PW-ILA DPD approach employs LS fitting to identify the post-inverse filter coefficients in a per-region basis, as in the reference paper \cite{PW_chalmers}. Hence, $K$ independent LS fits are calculated per ILA iteration. {The complexity of a complex block LS is $4(\ceil*{\frac{N_{\text{PW}}}{K}}^2\ceil*{\frac{B_{\text{ILA}}}{K}}+\ceil*{\frac{N_{\text{PW}}}{3K}})$ \cite{Matrix_computations}}.
\end{itemize}

While the complexity results are provided in Table~\ref{tab:complexity} in symbolic form, for generality, concrete numerical complexity values and comparisons are provided, along the RF measurement results, in Section \ref{Measurements}.

\begin{figure}[t!]
    \centering
    \begin{subfigure}[t]{1\linewidth }
        \includegraphics[width=\linewidth]{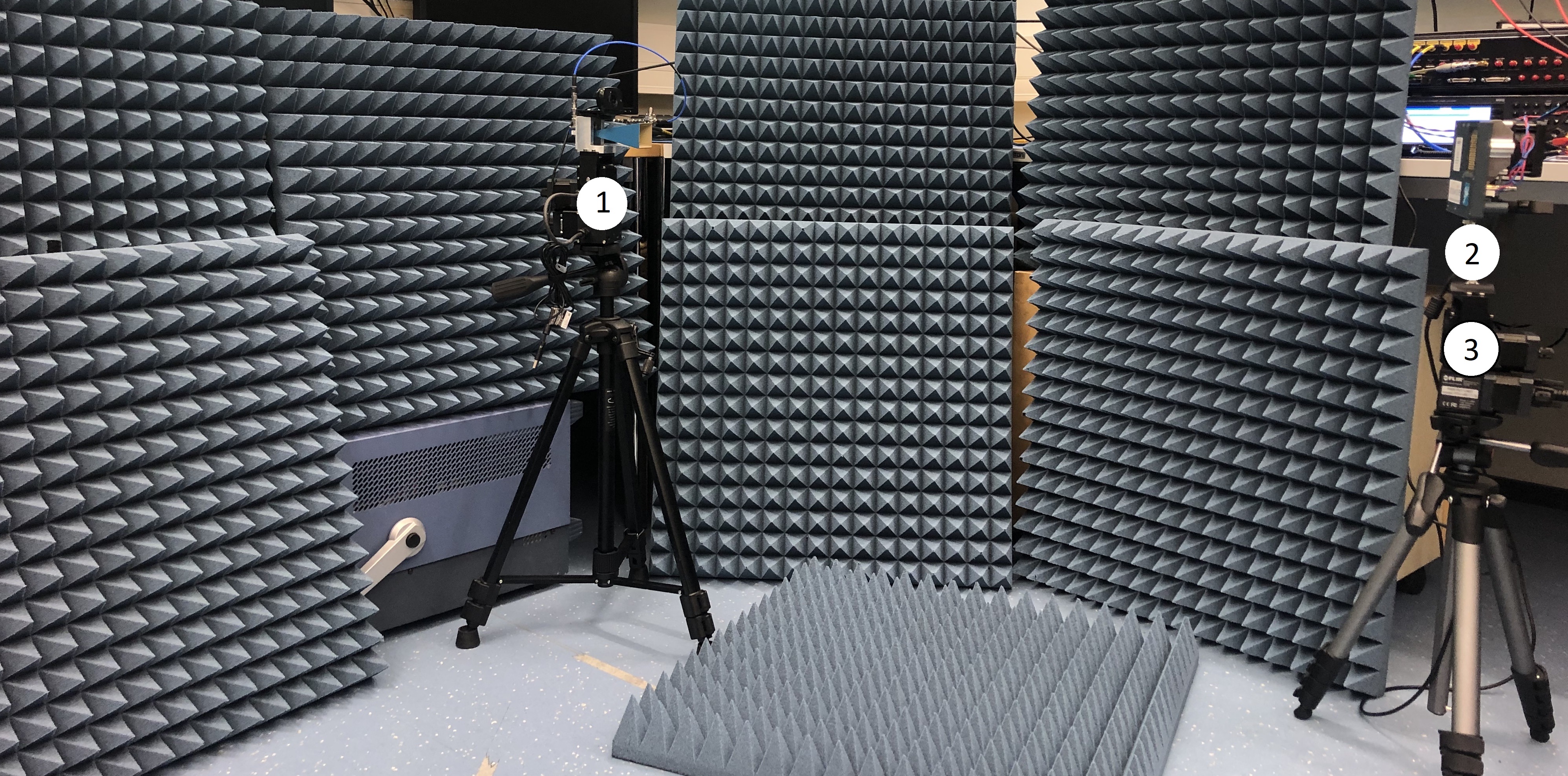}
    \end{subfigure}
    \begin{subfigure}[t]{1\linewidth}
        \includegraphics[width=\linewidth]{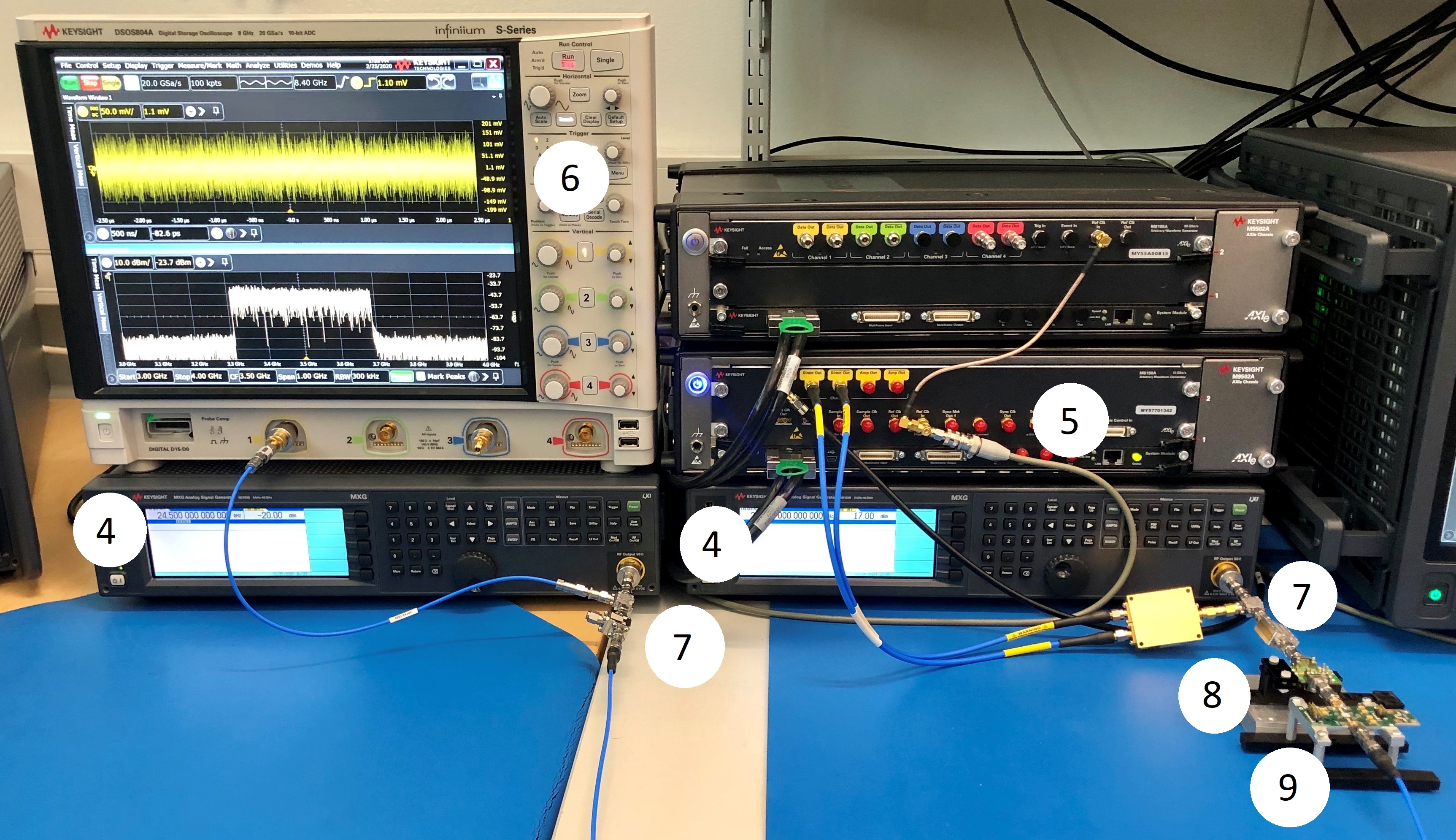}
    \end{subfigure}
    \vspace{-2mm}
    \caption{\quad OTA mmWave measurement setup at 28~GHz, {with individual elements being labeled as (1) observation horn antenna, (2) AWMF-0129, (3) mechanical rotator, (4) N5183B-MXG, (5) M8190A, (6) DSOS804A, (7) T3-1040, (8) HMC499LC4, and (9) HMC1131. Separation between the TX and RX antenna systems is 1.5 meters.}} \label{fig:setup}
\end{figure}

\vspace{20mm}

\section{mmWave Over-the-Air Measurements} \label{Measurements}
In this section, we provide extensive RF measurement results on a state-of-the-art 64-element (${8\times8}$) active antenna array operating at 28 GHz carrier frequency in order to demonstrate and evaluate the operation of the proposed DPD solution in the context of 5G NR systems at FR2.

\subsection{mmWave Measurement Setup}
The measurement setup is shown in Fig. \ref{fig:setup}. The Keysight M8190A arbitrary waveform generator is utilized to generate the TX IF signal centered at 3.5 GHz. Two Keysight N5183B-MXG signal generators running at 24.5 GHz are used to generate the local oscillator signals  that, together with two Marki Microwave mixers (T3-1040), are utilized for up-converting the IF signal to the desired carrier frequency, at TX, and for downconverting the high-frequency signal back to IF, at RX. The modulated RF waveform at 28 GHz is amplified with one HMC499LC4 and one HMC1131 driver amplifiers that allow to feed the Anokiwave AWMF-0129 active antenna array such that its PAs are driven close to saturation, yielding up to +44 dBm EIRP. The transmit signal then propagates over-the-air and is captured by a horn-antenna {located 1.5 meters} apart. After downconversion to IF, the signal is captured by the Keysight DSOS804A oscilloscope that is utilized as the observation and measurement receiver/digiter, and taken to baseband. The received samples are then processed in a host PC running \textsc{Matlab}, where the actual DPD learning and predistortion are performed.  Unless stated otherwise, the DPD is trained with the beam pointing towards 0 degrees direction.
Furthermore, in these measurement experiments, we utilize the actual OTA received signal to learn the DPD filter coefficients, as the utilized active array does not facilitate hardware based observation. {Additionally, different transmit signal realizations are always used, in DPD learning phase and in assessing the final DPD linearization performance.}

In all the measurement examples, a 400 MHz OFDM waveform with 64-QAM subcarrier modulation conforming to the 3GPP 5G NR downlink specifications \cite{3GPPTS38104}, with OFDM subcarrier spacing of 120 kHz, $K_{\text{ACT}}=3168$ active subcarriers, FFT size of $K_{\text{FFT}}=4096$ and with an oversampling factor of 5 is adopted. {The waveform generation also includes WOLA processing to improve the bandlimitation of the digital transmit waveform. The complementary CCDF of the sample-level PAPR of the PA input signal, after iterative clipping and filtering, reads 6.4~dB and 7~dB when measured at $1\%$ and $0.01\%$ points, respectively}. {This imposes an EVM limit of some $4\%$ to the digital transmit waveform. Under these test conditions, the output 1~dB compression point of the Anokiwave AWMF-0129 was measured to correspond to an EIRP of ca. +42~dBm}. {Additionally, it is noted that no amplitude control is considered in the analog beamformer.}

The proposed PW-CL solution employs $I_{\text{CL}} = 10$ block iterations of $B_{\text{CL}} = 20.000$ samples each. As benchmark, we consider the single-polynomial CL DPD \cite{Concurrent_abdelaziz,DPD_MM_5} with the same number of block iterations and samples, as well as the PW-ILA DPD \cite{PW_chalmers} and classical single-polynomial ILA, both of them consisting of $I_{\text{ILA}} = 4$ ILA iterations of $B_{\text{ILA}} = 50.000$ samples each. For the region partition we consider $Q_k=5$ and $e_k =0.01$ for each region, resulting in $K=3$ regions, while the same amount of regions is also used in the reference methods. All the nonlinear orders and memory depths of the model in (\ref{eq:dual_input_model_final}) are set to $9$ and $3$, respectively. 

{Finally, it is noted that for the purpose of reliably assessing the linearization performance, in terms of, e.g., the linearized spectrum and the ACLR metrics, noise averaging is utilized to reduce the noise floor of the OTA measurement receiver. This way, enhanced dynamic range of around 54~dB can be achieved, allowing thus for reliable quantitative performance measurements of different considered linearization solutions. For reference, the relative noise floor of the measurement receiver system is also depicted along Fig. \ref{fig:beam_pattern}.}

\begin{figure*}[t!]
    \centering
    \begin{subfigure}[t]{0.4\textwidth }
        \includegraphics[width=\textwidth]{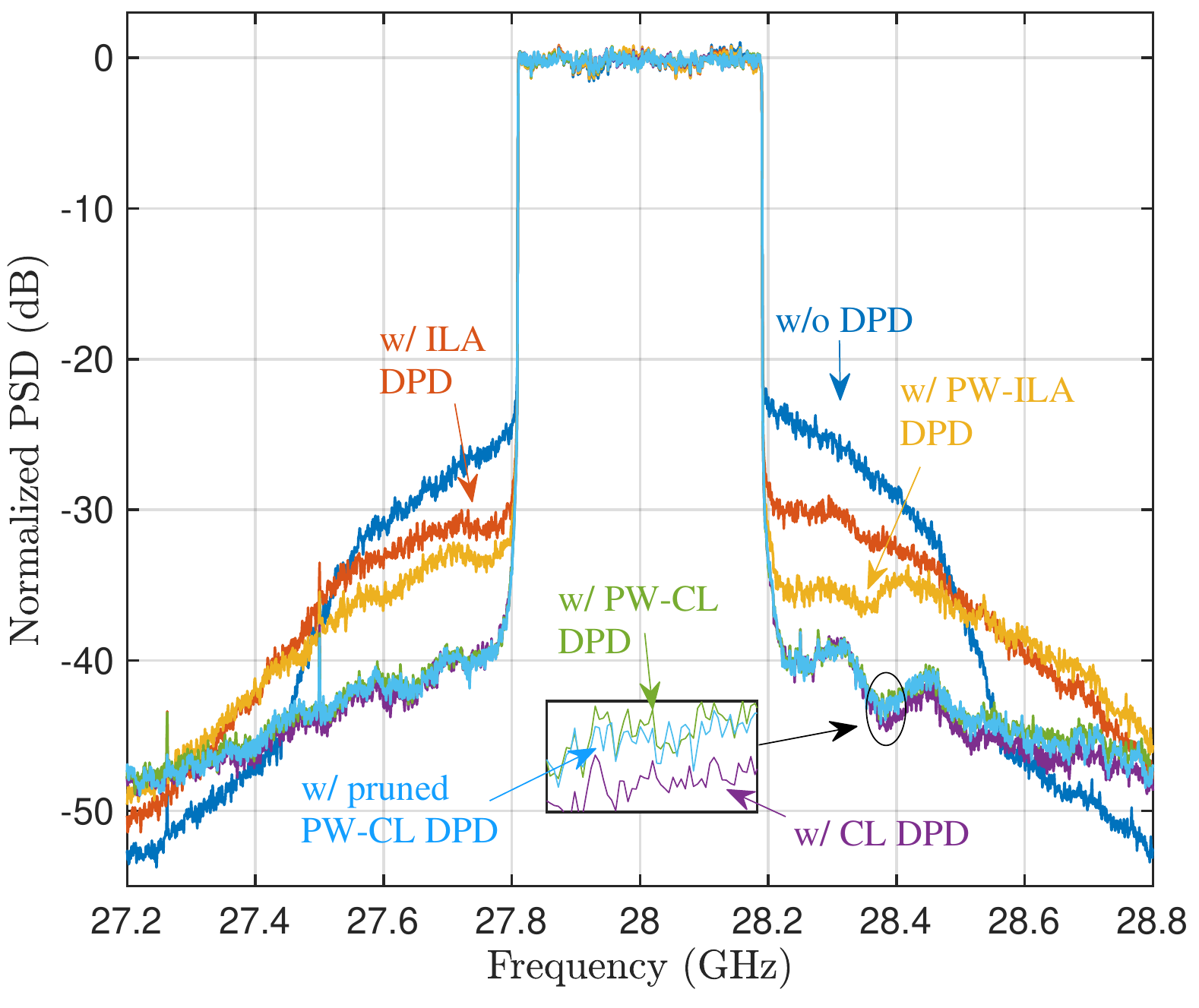}
        \caption{}
        \label{Fig:medium_power}
    \end{subfigure}
    \begin{subfigure}[t]{0.4\textwidth}
        \includegraphics[width=\textwidth]{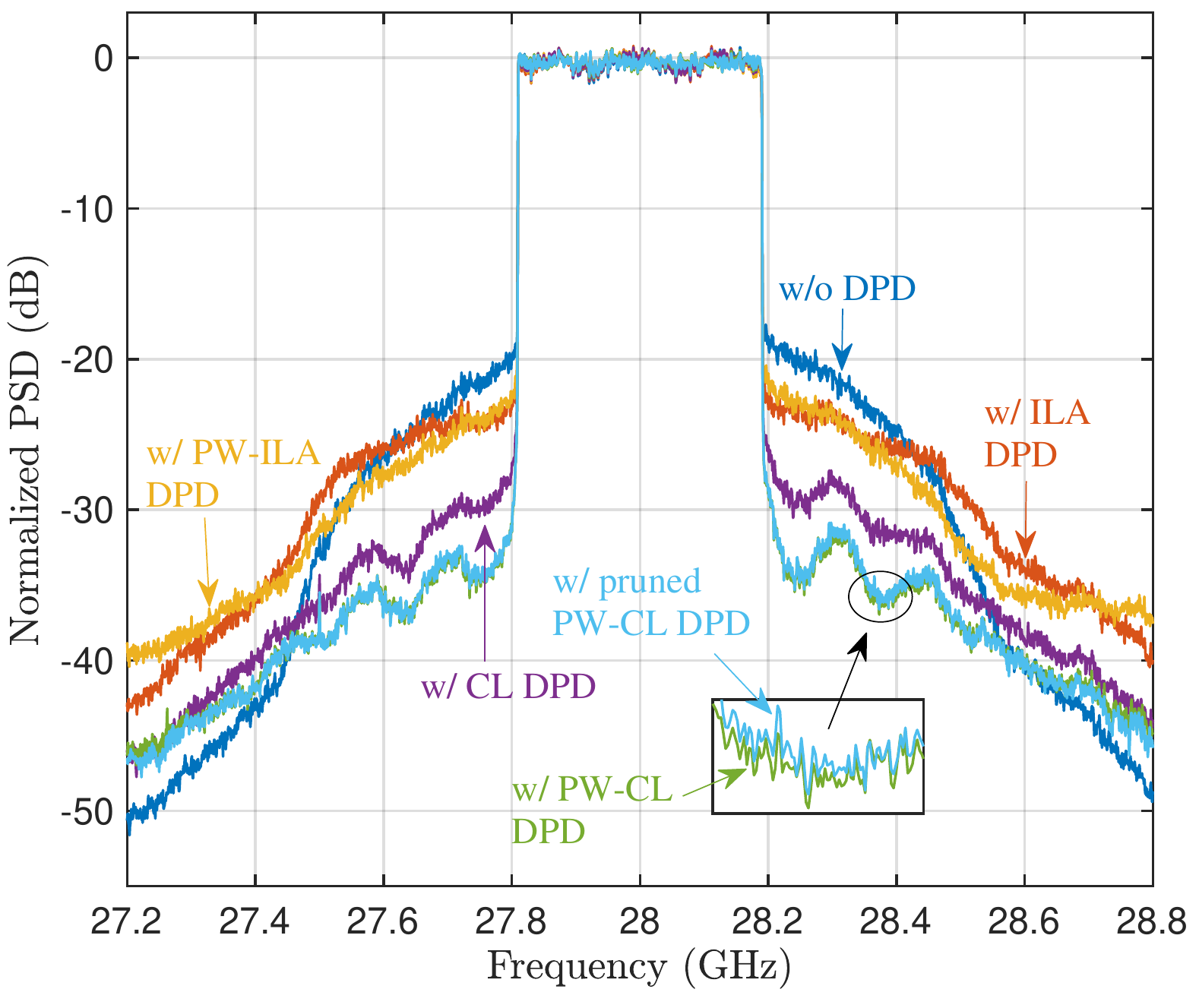}
        \caption{}
        \label{Fig:high_power}
    \end{subfigure}
    \vspace{-2mm}
    \caption{\quad  Example OTA observed spectra for the proposed PW-CL DPD and reference DPD solutions, for (a) \text{EIRP} of $ +41.3 \text{dBm}$, and for (b) \text{EIRP} $ +43.3 \text{dBm}$.} \label{fig:figure_1}
\end{figure*}

\begin{table}[t!]
\caption{\textsc{5G NR Release-15 EVM Requirements at FR2 \cite{3GPPTS38104}}}
\centering
{\begin{tabular}{lcc}\hline\vspace{-1.7mm}  & & \\\vspace{1mm}
 Modulation          &EVM ($\%$)  & EVM (dB)    \\\hline & & \vspace{-2mm} \\  
16-QAM                      & 12.5        & -18       \\
64-QAM 				  & 8        & -22       \\ & & \vspace{-2.75mm}\\ \hline
\end{tabular}}{}
\label{tab:EVM}
\end{table}

\begin{table}[t!]
\caption{\textsc{5G NR Release-15 ACLR Requirements at FR2 \cite{3GPPTS38104}}}
\centering
{\begin{tabular}{lc}\hline\vspace{-1.7mm}  &\\\vspace{1mm}
Frequency band          & ACLR ($\text{dBc}$)     \\\hline & \vspace{-2mm} \\
24.25 - 33.4 (GHz)                      &28               \\
37 - 52.6 (GHz) 				  &26             \\ & \vspace{-2.75mm} \\\hline
\end{tabular}}{}
\label{tab:ACLR}
\end{table}

\subsection{5G NR Waveform Quality Requirements at FR2}
Due to the high level of integration of active antenna array systems considered at FR2, where the front-ends are built in the actual antenna array, access to the antenna ports and running the corresponding conducted conformance tests are no longer feasible. Consequently, 3GPP has defined new methods to characterize the {transmitter system} EVM and ACLR by means of radiated conformance testing, which are described in \cite{3GPPTS38104} and \cite{3GPPTS38141-2}, {and adopted recently also in \cite{Tervo_OTA}}.  
The EVM is measured for the effectively OTA combined signal by using a measurement receiver, and is defined as follows 
\begin{equation}
\text{EVM}_{\%} = \sqrt{P_{\text{error}}/P_{\text{ref}}} \times 100\%,
\end{equation}
where $P_{\text{error}}$ is the power of the error signal defined as the difference between the ideal transmit symbols and the corresponding complex samples at the measurement receiver, after amplitude and phase equalization, while $P_{\text{ref}}$ is the average power of the ideal constellation symbols. The specified EVM requirements for different currently supported modulation schemes are gathered in Table \ref{tab:EVM}.

The ACLR is also characterized through OTA measurements and is defined as the ratio of the TRP centred on the assigned frequency to the filtered mean TRP centred on an adjacent channel frequency, formally given as
\begin{equation}
    \text{ACLR}= 10\log_{10}\frac{\text{TRP}_{\text{channel}}}{\text{TRP}_{\text{adjacent}}}, \label{eq:aclr}
\end{equation}
where $\text{TRP}_{\text{channel}}$ denotes the TRP within the assigned channel, and $\text{TRP}_{\text{adjacent}}$ is the maximum TRP of the two adjacent channels. The measurement bandwidth is defined as the bandwidth containing $99\%$ of the radiated allocated channel power in the direction given by $\theta$ and $\phi$. The adjacent channel measurement bandwidth is equal to this. The TRP itself is defined as \cite{3GPPTS38104}
\begin{equation}
\begin{split}
    \text{TRP} &\approx \frac{\pi}{2AE}\\
    &\times \sum_{\substack{n=0}}^{A-1} \sum_{\substack{m=0}}^{E-1}\left(\text{EIRP}_{p1}(\theta_n,\phi_m) + \text{EIRP}_{p2}(\theta_n,\phi_m)\right)\text{sin}\theta_m,
\end{split}\label{eq:TRP}
\end{equation}
where $A$ and $E$ refer to the total number of samples in elevation {($\theta$)} and azimuth {($\phi$)}, respectively, while $\text{EIRP}_{p1}$ and $\text{EIRP}_{p2}$ refer to the EIRP in two orthogonal polarizations $p_1$ and $p_2$, respectively. 
The ACLR requirements for FR2 are gathered in Table \ref{tab:ACLR}.

\subsection{{Baseline OTA Measurement Results and Complexity Assessment}}

{Next, linearization performance results and the corresponding complexity numbers and comparisons are provided, at two example EIRP levels. In case of the proposed PW-CL DPD system, the linearization performance is assessed by using the learning rule in (\ref{eq:learning_rule2}) along with orthogonal and pruned BFs, while the complexity evaluations are provided also for the learning rule in (\ref{eq:learning_rule1}) with non-orthogonal BFs. Later, in Section \ref{sec:power_sweep}, the linearization performance with the learning rule in (\ref{eq:learning_rule1}) is also demonstrated, and shown to be essentially identical to that obtained through (\ref{eq:learning_rule2}). }

\begin{figure}[t!]
\centering
\includegraphics[width=.9\linewidth]{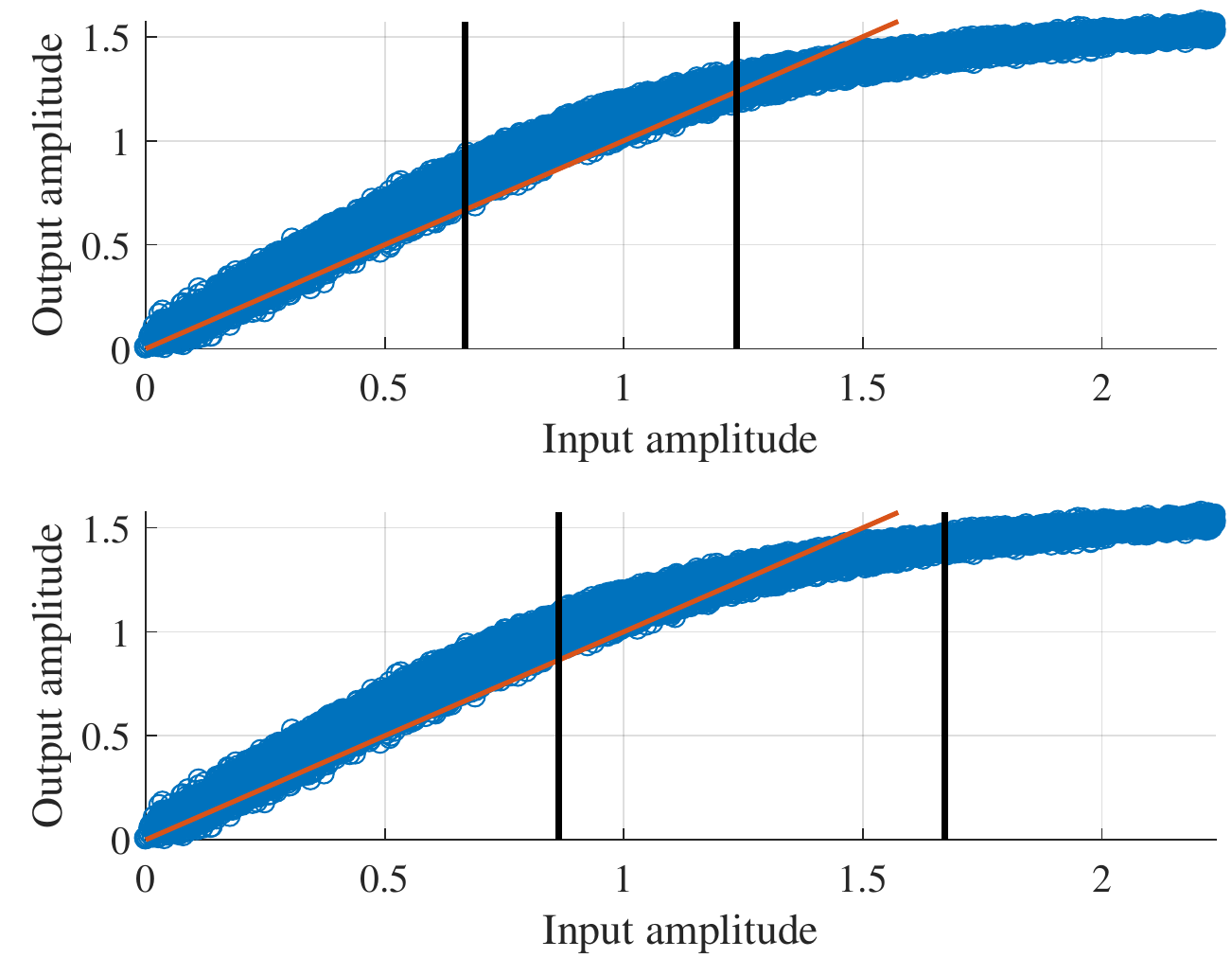}
\caption{\quad Region partition provided by \cite{PW_chalmers} (top), and by Algorithm \ref{alg:Partition_algorithm} (bottom) for an EIRP of +43.3~dBm. Reference red line represents the ideal linear response.}
\label{fig:region_partition_array}
\end{figure}

\begin{figure}[t!]
\centering
\includegraphics[width=.85\linewidth]{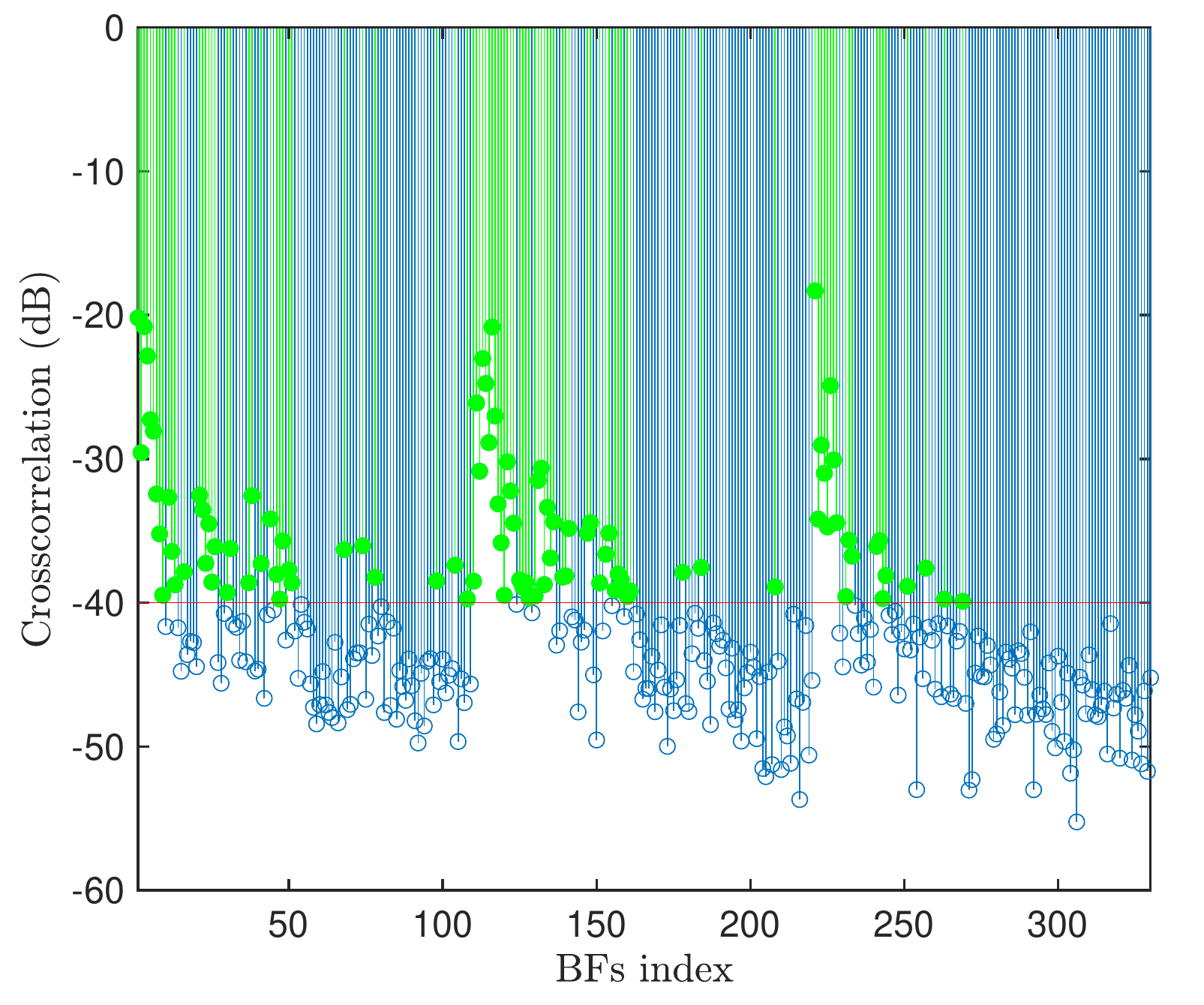}
\caption{\quad Illustration of basis function pruning. In blue, the initial set of BFs and in green, the ones selected by the pruning algorithm for an EIRP of +43.3~dBm.}
\label{fig:pruned_BFs}
\end{figure}

 Fig. \ref{fig:figure_1} illustrates the spectra of the OTA observed signals corresponding to two different EIRP levels, when the proposed DPD solutions along with the reference methods are considered. As it can be observed, both ILA-based methods clearly underperform, while also the classical CL DPD solution deteriorates at the higher TX power case. On the other hand, the proposed PW-CL DPD solution provides excellent linearization at both power levels. The regions for PW processing provided by the $K$-means reference algorithm and the proposed algorithm are shown in Fig. \ref{fig:region_partition_array}, where it can be observed that the $K$-means algorithm tends to gather the regions around the average amplitude value, while the proposed solution provides more evenly distributed regions. In this experiment, we have also considered to prune the orthogonalized BFs of the proposed PW-CL solution, such that the threshold is set to $\zeta_{\text{TH}} = -50$~dB and $-40$~dB for the scenarios considered in Fig. \ref{fig:figure_1}(a) and Fig. \ref{fig:figure_1}(b), respectively. Thus, the BFs whose corresponding $\boldsymbol{\zeta(i)}$ are above this threshold are utilized to learn and predistort the transmit signal, as depicted in Fig. \ref{fig:pruned_BFs} for the scenario of Fig. \ref{fig:figure_1}(b). The pruning algorithm allows to reduce the number of BFs from 348 to 96 while keeping similar linearization performance. Further pruning can be achieved by increasing the threshold, but at the expense of reduced linearization. For instance, the threshold could be optimized to fulfil the EVM and ACLR values in Tables \ref{tab:EVM} and \ref{tab:ACLR}, thus minimizing the number of coefficients.
 
\begin{table*}[]\centering \caption{{\textsc{DPD main path processing complexity and parameter learning complexity, both expressed as FLOPs/sample, \\ for the linearization experiment at EIRP of +43.3}~dBm.} 
}\label{tab:complexity_experiment}
\begin{tabular}{cccccccccc}\hline \vspace{-1.8mm} & & & & & & & & \\
                  &  & \multirow{2}{*}{\vspace{-2mm}\textbf{PW-ILA}} &  & \multicolumn{2}{c}{\textbf{CL}} &  & \multicolumn{3}{c}{\textbf{Proposed PW-CL}} \vspace{0.5mm}  \\  \cline{5-6} \cline{8-10} \vspace{-1.8mm} & & & & & & & & \\
                  &  &                   &  &      Self orth.     &    Orth. BFs      &  &   Self orth.   &    Orth. BFs    &   Orth and pruned BFs  \vspace{0.5mm} \\  \clineB{1-10}{2.5} & & & & & & & & \vspace{-2mm}\\ 
\multirow{4}{*}{\makecell{\textbf{DPD} \\ (FLOPs/sample)}} & BF gen. &    $9$               &  &  $9$         &    $9$      &  &  $9$     &   $9$    &  $9$  \vspace{0.7mm}  \\ 
                  & Filt.  &    $926$               &  &     $926$      &     $926$     &  &  $926$     &   $926$    &   $249$  \vspace{0.7mm} \\
                  & Orth. &     $-$              &  &     $-$      &      $26,912$    &  &  $-$     &   $26,912$    &   $1,964$  \vspace{0.7mm} \\
                  & Total  &           $935$        &  &     $935$      &     $27,847$     &  &  $935$     &   $27,847$    &   $2,219$   \\ & & & & & & & &\\
\multirow{3}{*}{\makecell{\textbf{Learning} \\ (FLOPs/sample)}} & BF gen. & $9$                  &  &    $-$       &   $-$       &  &   $-$    &    $-$   &   $-$ \vspace{0.7mm} \\
                  & DPD est. &     $2.7\times10^9$              &  &     $54,520$      &    $928$      &  &    $54,520$   &  $928$     &  $323.2$  \vspace{0.7mm}  \\
                  & Total &          $2.7\times10^9$          &  &      $54,520$     & $928$         &  &   $54,520$    &    $928$   &   $323.2$  \vspace{0.7mm} \\ \hline
\end{tabular}
\end{table*}

{The exact complexity numbers of the different DPD methods corresponding to the experiment at EIRP of +43.3~dBm are gathered in Table \ref{tab:complexity_experiment}, building on the general complexity expressions in Table \ref{tab:complexity}, and considering the number of learning block iterations, learning block-size, and DPD parameterization reported at the beginning of this subsection. {It is noted that the numerical learning complexities have been normalized by the total number of training samples, so that they are also expressed in terms of FLOPs/sample. This is done in order to be able to do relative complexity comparisons between the main path linearization and parameter learning in more straight-forward manner}. As it can be observed, the CL learning solutions entail much lower learning complexity than ILA, in general. {Utilizing orthogonal and pruned BFs allows to significantly reduce the learning complexity -- at the expense of a slight increase of the main-path complexity -- which may facilitate real-time DPD  adaptation if needed.} Overall, this experiment shows the superiority of the proposed PW-CL DPD method compared to current state-of-the-art, both in terms of linearization performance and complexity. {Specifically, depending on whether the main path or the learning complexity is emphasized, the proposed PW-CL DPD system should be executed with non-orthogonalized BFs and self-orthogonalized learning or with orthogonalized and pruned BFs and classical LMS-like learning, respectively.}}

In the following additional experiments, we study the performance of the PW-CL DPD in further details, while do not anymore extensively compare against all the previous reference methods. The proposed solution is, however, still further benchmarked against the PW-ILA solution in \cite{PW_chalmers}, which stands as an improved version of the single-polynomial ILA-based method that is widely adopted in current art \cite{Full_angleDPD,reduced_set_DPD,Our_OTA_DPD, OTA_DPD1,OTA_DPD2,OTA_DPD3}.

\begin{figure}[t!]
\centering
\includegraphics[width=.9\linewidth]{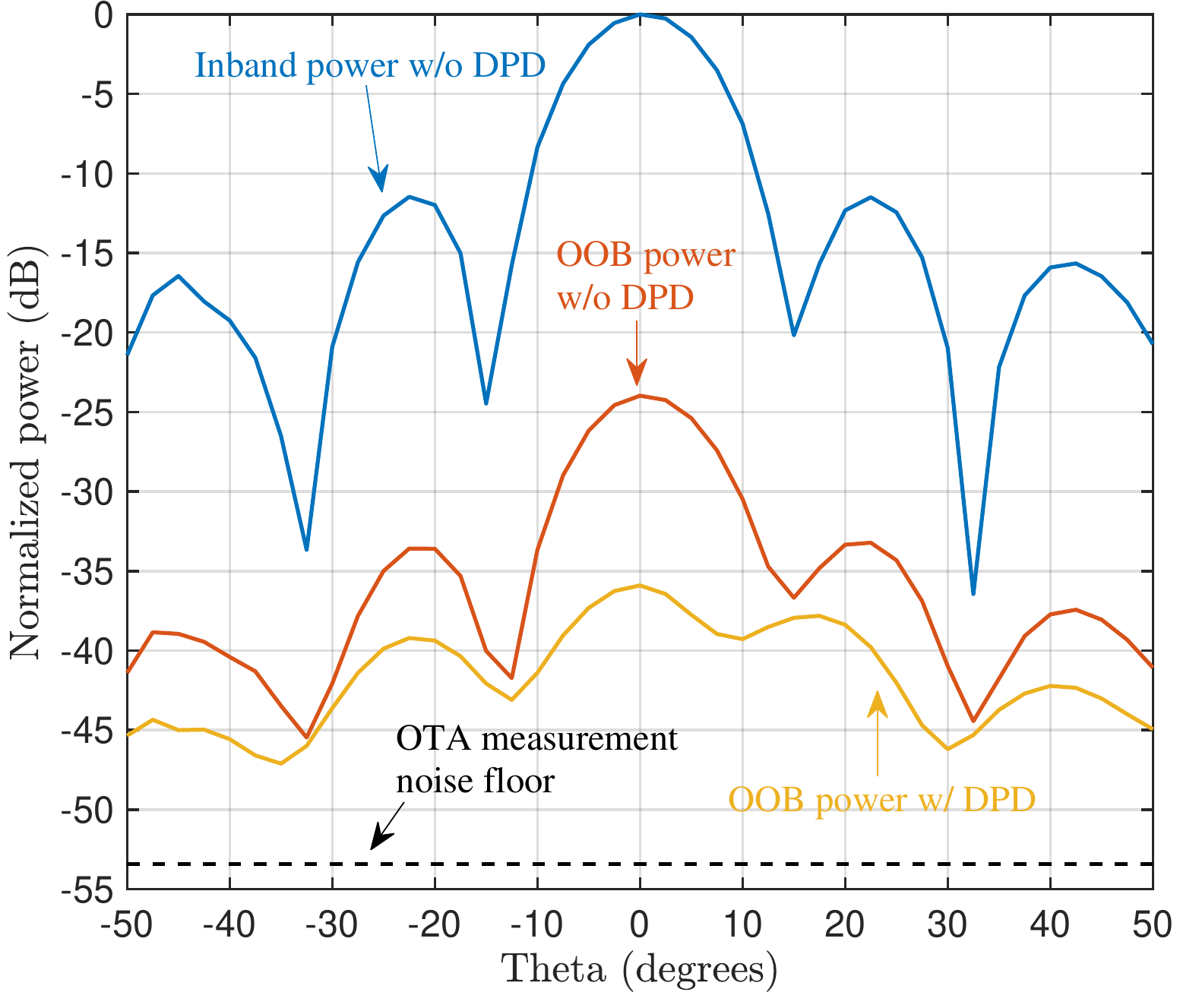}
\caption{\quad Measured normalized far-field beam-patterns of the in-band power and OOB emissions, at EIRP of +43.3 dBm.}
\label{fig:beam_pattern}
\end{figure}

\begin{figure*}[t!]
    \centering
    \begin{subfigure}[t!]{0.4\textwidth }
        \includegraphics[width=\textwidth]{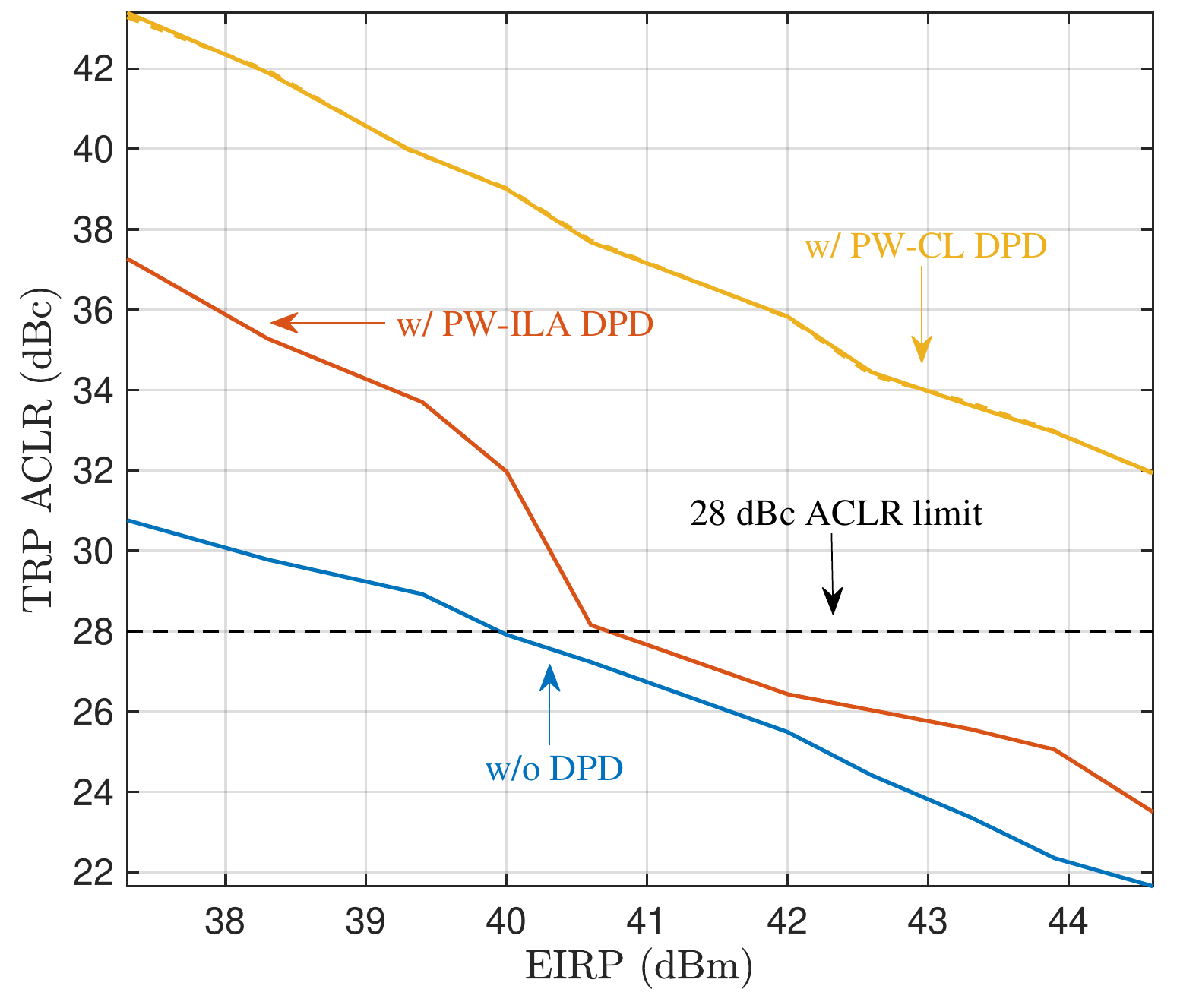}
        \caption{}
        \label{Fig:ACLR_power_sweep}
    \end{subfigure}
    \begin{subfigure}[t!]{0.4\textwidth}
        \includegraphics[width=\textwidth]{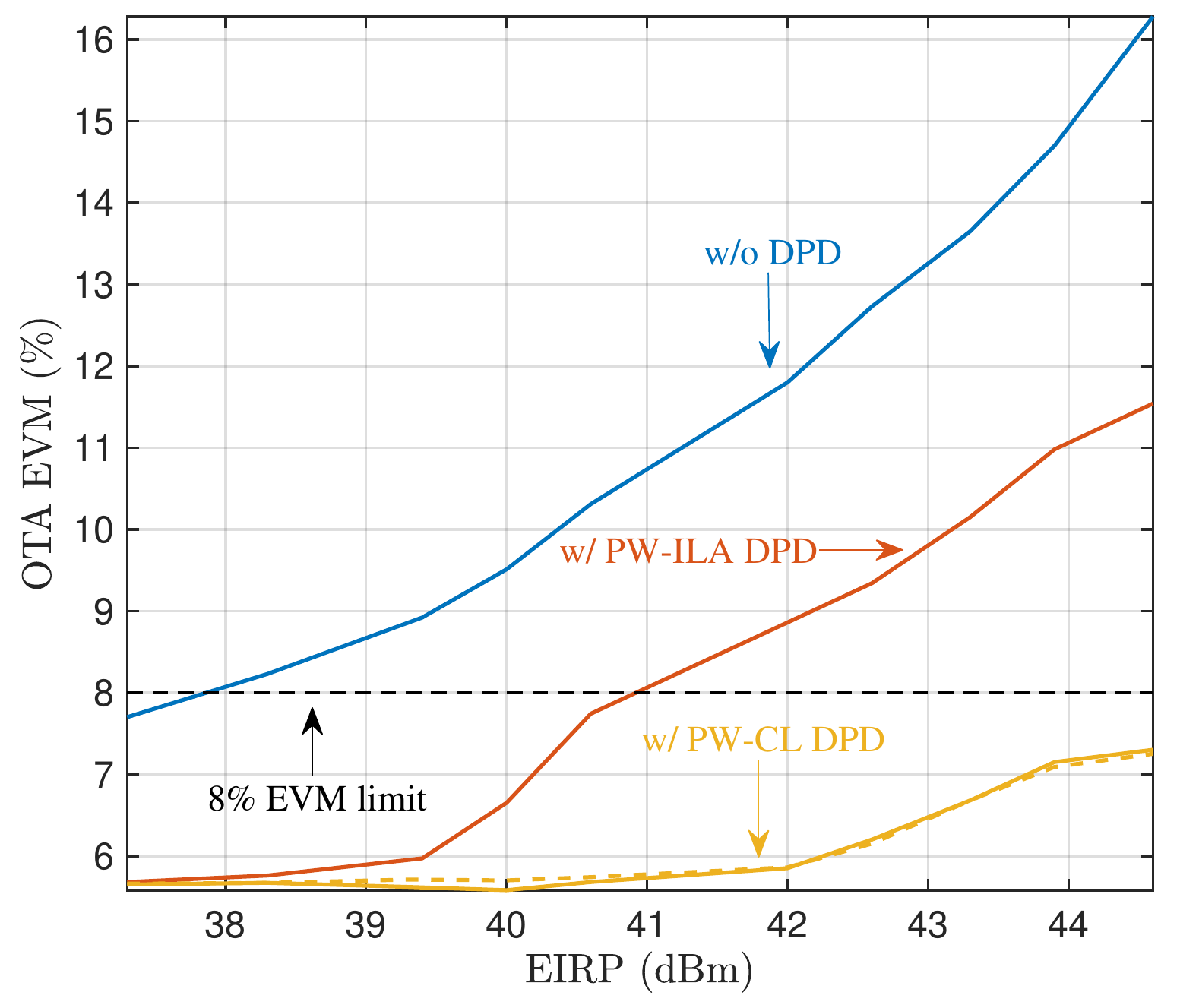}
        \caption{}
        \label{Fig:EVM_power_sweep}
    \end{subfigure}
    \vspace{-2mm}
    \caption{\quad {Linearization performance in terms of (a) ACLR and (b) EVM as a function of the EIRP for the proposed PW-CL DPD either employing orthogonal BFs and the learning rule in (\ref{eq:learning_rule2}), shown in yellow solid lines, or employing the self-orthogonalized learning rule in (\ref{eq:learning_rule1}) and non-orthogonal BFs, shown in dashed yellow lines. Also the performance without DPD and with reference PW-ILA DPD \cite{PW_chalmers} are shown.}}\label{fig:figure_2}
\end{figure*}

\subsection{Power-Sweep OTA Measurements}\label{sec:power_sweep}
{In the next experiments, the linearization performance of the PW-ILA DPD and the PW-CL DPD, employing both learning rules in (\ref{eq:learning_rule1}) and (\ref{eq:learning_rule2}), are further analyzed, in terms of ACLR and EVM, as a function of the EIRP level}. This allows to analyze how much additional power the active array can deliver when different DPD solutions are considered while meeting the different FoM, and how the performance decays as the array is operated deeper and deeper into compression.

In order to evaluate ACLR based on the TRP metric defined in (\ref{eq:TRP}), the received power levels in the inband and OOB regions need to be measured for many angular directions.
In order to conduct such a measurement, the antenna array, mounted on a digitally-controlled mechanical rotator, is rotated from -50 to 50 degrees with an angular resolution of 2 degrees, {which fulfills the angular step criteria specified by 3GPP in \cite{3GPPTS38141-2}}. {For the considered array system and carrier frequency, the angular resolution should be at least ca 8.68 degrees}. Then, the horn antenna at a fixed location measures the received power for each angle. This effectively yields the far-field beam-pattern of the array, as depicted in Fig. \ref{fig:beam_pattern} for the scenario shown in Fig. \ref{fig:figure_1}(b), illustrating the inband and OOB beam-patterns with and without DPD. {The inband and OOB powers are calculated from the OTA received signal after FFT processing by integrating over the corresponding frequency bins. For the OOB powers, only the worst-case adjacent channel is considered}. From this figure, we can make two important observations. First, the largest linearization is achieved in the mainbeam direction, which is a direct consequence of considering the main beam signal for DPD learning. Second, linearization is provided for the rest of the angular directions as well, although less than along the main beam. Consequently, the joint effect of the beam-pattern and the DPD keeps the OOB emissions low in all angular directions, {at least with this particular array HW.}

To calculate the ACLR numbers, the measured received powers are integrated over the measured angular directions, separately for inband and OOB regions, and then the ratio of the inband and the larger OOB TRP values is taken to calculate the ACLR as in (\ref{eq:aclr}). {It is noted that due to the constraints on the measurement setup, the elevation angle is fixed while only the azimuth angle is varied when evaluating the TRP metric.}
It is also noted that previous works in array linearization \cite{DPD_DigitalMIMO,DPD_MM_4,DPD_MM_5,DPD_MM_6,OTA_combining_DPD,Full_angleDPD,reduced_set_DPD,Our_OTA_DPD,OTA_DPD1,OTA_DPD2,OTA_DPD3} have calculated ACLR through the ratio between the inband and the adjacent powers in the main-beam direction only. By doing so, while considering only the main-beam direction for DPD learning, the calculated ACLR is commonly around 5~dB better than that calculated through TRP for the example shown in Fig. \ref{fig:beam_pattern}, which is very far from its actual value. The difference essentially stems from the fact that the beam-patterns without DPD are well defined, and most of the energy is being transmitted towards the main-beam direction, which dominates the TRP metric. However, the OOB beam-pattern with DPD has comparable power levels for a wider range of angles and all of them contribute significantly to the corresponding TRP. Thus, considering only the main-beam for ACLR evaluation yields highly optimistic values.  

The power sweep measurements are gathered in Fig. \ref{fig:figure_2}. In order to assess how much higher TX power or EIRP the different DPD methods allow to obtain, it is necessary to compare for which EIRP levels either the EVM or ACLR metrics are no longer fulfilled. For the no-DPD reference case, the first FoM not being fulfilled is the EVM for an EIRP of around +38~dBm, while for the PW-ILA DPD it is the ACLR at an EIRP of +40.6~dBm. The proposed PW-CL DPD allows to fulfill both metrics across the considered range of EIRP levels, {regardless of the adopted learning rule}, being capable of providing more than 6 dB of extra EIRP compared to the no-DPD scenario and more than 4 dB compared to the PW-ILA DPD case. This will allow, e.g., to reduce the number of transmit antennas needed to deliver a desired target EIRP level, increase the coverage area, or improve the link budget. 

\begin{figure*}[t!]
    \centering
    \begin{subfigure}[t]{0.4\textwidth }
        \includegraphics[width=\textwidth]{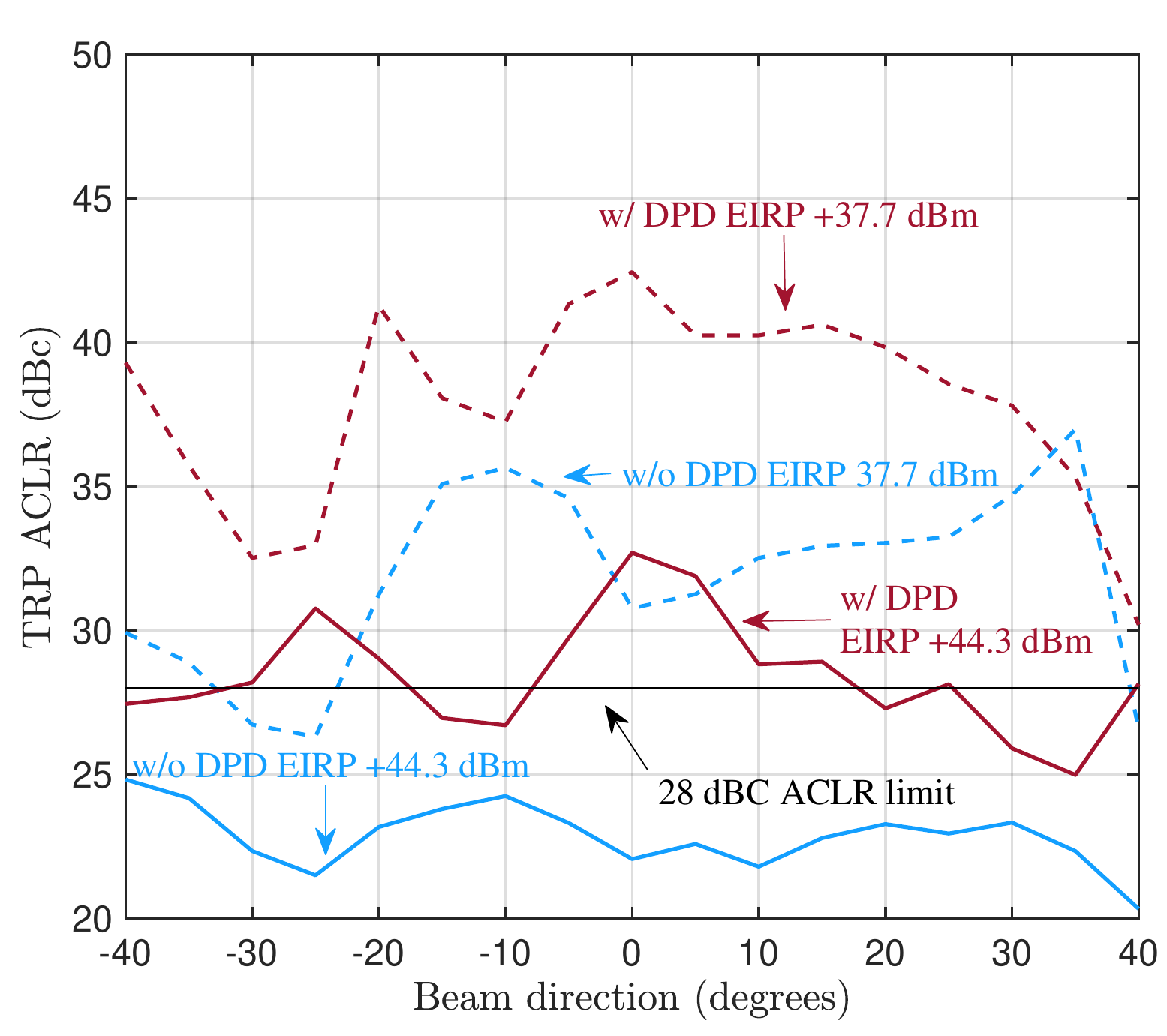}
        \caption{}
    \end{subfigure}
    \begin{subfigure}[t]{0.4\textwidth}
        \includegraphics[width=\textwidth]{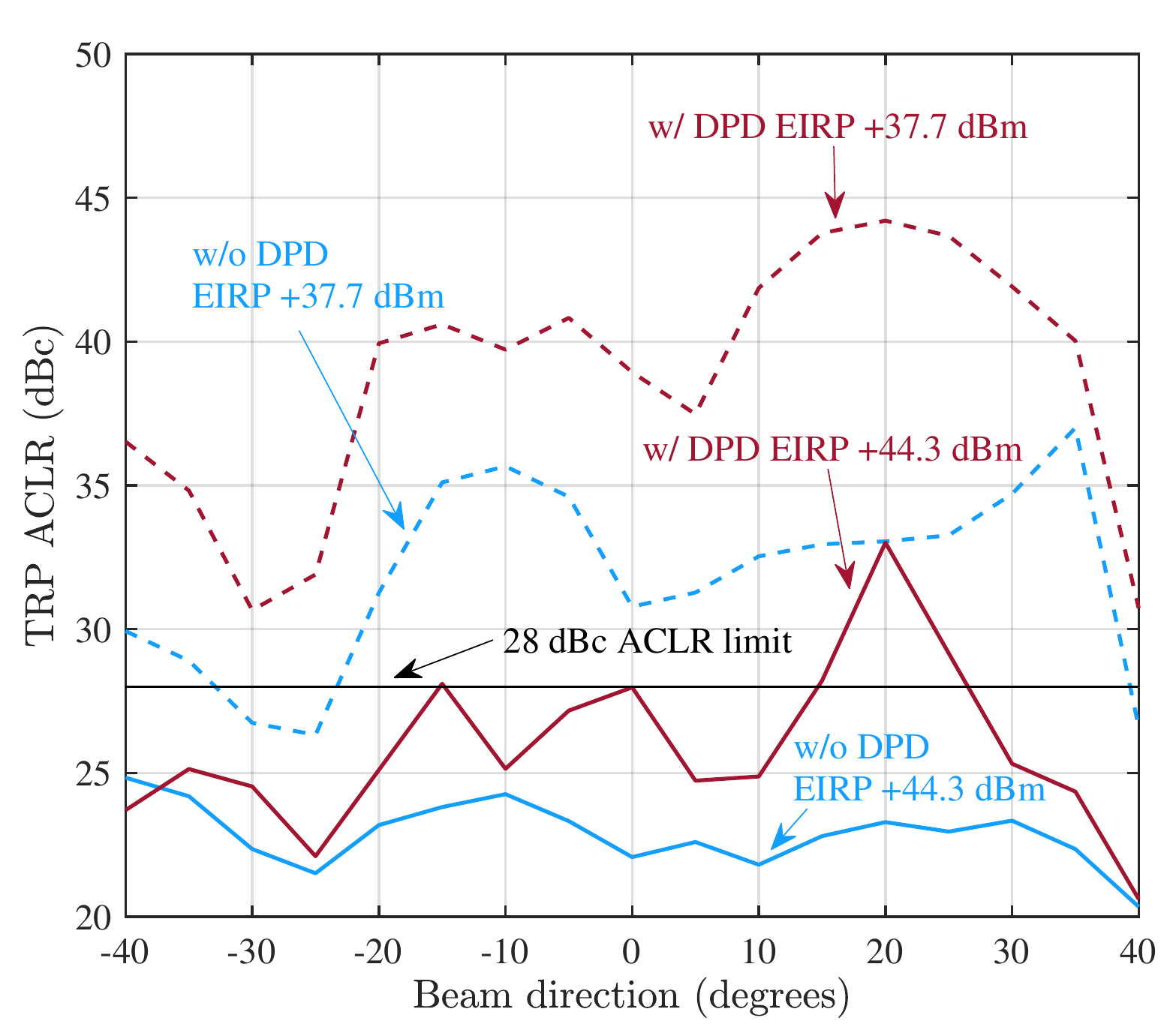}
        \caption{}
    \end{subfigure}
    \vspace{-2mm}
    \caption{\quad Linearization performance as the function of the steering angle when the DPD is trained with the beam pointing towards (a) 0 degrees direction and kept fixed, {and (b) 20 degrees direction and kept fixed.}}\label{fig:beam_dependency}
\end{figure*}

\subsection{Beam-Dependence of Radiated Nonlinear Distortion}
In this section, we investigate the effect of beamsteering on the nonlinear characteristics of the active antenna array. In current state-of-the art, this issue is generally overlooked, especially when considering actual mmWave active arrays with large number of antenna and PA units and considerable TX power or EIRP values. With this last experiment we try to shed some light on the importance of considering the beam-dependency of the load modulation when developing DPD methods tailored for active antenna arrays. 

The experiment was conducted as follows. The receiving antenna was located at zero degrees off the normal of the array, and the DPD coefficients were learnt considering the main beam pointing towards this direction -- {so far identical to what was done also in the previous experiments}. The position of the receiving antenna and the DPD coefficients were kept fixed throughout the whole experiment. Then, the electrical beam was steered from -50 to 50 degrees with a resolution of 5 degrees {by means of digitally controlled phase-only analog beamforming}. For every beam direction, the antenna array was mechanically rotated towards the opposite direction by the same amount, so that the main beam is always perfectly aligned with the receiver antenna. Lastly, the ACLR was measured through the TRP as defined in Section \ref{sec:power_sweep}, {for the worst-case adjacent channel}. {The same experiment was then also repeated but such that the beam was pointed towards 20 degrees direction when training the DPD.}  {The results of both experiments are depicted in Fig. \ref{fig:beam_dependency} (a) and (b), respectively}, for two considered transmit power scenarios. As it can be observed, excellent linearization is achieved at those directions at which the DPD was trained. However, as the beam is swept, there is a systematic loss of linearization performance due to the load modulation modifying the exact nonlinear characteristics of the array. This phenomenon is more noticeable for the high transmit power case, which implies that the load modulation becomes more severe when the array is operated deeper in compression. For the highest EIRP level, the DPD does not fulfill the ACLR requirements even over a 20 degree beam-tuning range.
On the other hand, applying some back-off alleviates this issue, and the ACLR limit is fulfilled over a wider range of angles, at the expense of reduced energy efficiency, output power and network coverage. {It is also interesting to observe that, in general, the DPD works relatively well around the angle at which it was trained, and the way it deteriorates depends mainly on how similar/different the nonlinear characteristics are compared to those at the considered training angle. Overall, these findings indicate that even continuous learning and adaptation of the DPD system may be needed in networks where fast beam-tuning is adopted. } 

\subsection{Discussion on Learning Approaches}
DPD learning becomes a very challenging task when the array exhibits beam-dependent nonlinear behavior. If energy efficiency and EIRP are to be maximized while meeting the FoM, consistent linearization performance across all steering angles and transmit powers needs to be delivered. One approach to ensure proper DPD operation is to execute the DPD learning near-continuously, such that it is possible for the DPD to keep track of fast changes in the nonlinear conditions due to beamsteering. To do so, access to the PA outputs is necessary to build the learning signal that characterizes the far-field combined signal, possibly requiring directional couplers to be implemented at the PA outputs, as well as the co-phasing and combining (anti-beamforming) unit. In a TDD system, it is in principle possible to utilize the receiver-side analog beamformers for anti-beamforming, thus alleviating implementation costs. A possible alternative to the on-line learning approach is to have a set of predetermined DPD coefficients stored for every beam direction or range of directions, as well as for different transmit powers, as discussed also in \cite{reduced_set_DPD}. However, adaptability to changes due to temperature or device aging, for example, would be limited. On the other hand, relying on far-field test receivers might be overly complicated as several test receivers would be required to capture the beam-dependency behavior. Furthermore, direct observation of the main beam signal with far-field receivers may not be possible, as the main beam is being pointed towards the intended users during the data transmission. It is also important to highlight that the actual crosstalk, and hence the severeness of the load modulation, depends on the considered transmitter hardware, and thus, its effects on the DPD performance should be investigated carefully for each practical scenario.

\section{Conclusions}\label{sec:Conclusions}

In this paper, a framework for effective linearization of strongly nonlinear active antenna array transmitters was presented and described, building on piecewise DPD processing and closed-loop gradient-descent parameter learning. 
For efficient PW modelling and processing, a region partition algorithm that takes into account the actual nonlinear characteristics of the device was also proposed. Additionally, in order to enable reduced complexity operation, a basis function pruning algorithm that leverages the information provided by the CL learning algorithm, was proposed and shown to significantly reduce the number of model coefficients without compromising its linearization capabilities. The proposed techniques were benchmarked against the current state-of-the-art, and tested and evaluated through extensive OTA RF measurements utilizing a 64-element active antenna array operating at 28 GHz carrier frequency. Through the ACLR evaluations, essentially conforming to the 5G NR conformance testing specifications, the proposed techniques were shown to consistently outperform the reference solutions, in particular when operating close to the saturation region. Specifically, it was shown that the proposed PW-CL DPD system with the new region partition method can provide more than 6 dB higher EIRP than the no-DPD reference system, and more than 4 dB higher EIRP when compared to the prior-art DPD techniques, {while still allowing for reduced total processing complexity}. The proposed methods thus allow for operating the active antenna array at a much more nonlinear and power-efficient operation point, while at the same time improving the mmWave network coverage. The load modulation phenomenon was also investigated, and shown to limit the angular range over which the DPD is effective, for given DPD coefficients. It was also shown that this phenomenon becomes more pronounced as the transmit power is increased, thus calling for frequent or even continuous updating of the DPD system -- {something that the gradient-adaptive CL systems can facilitate with feasible implementation complexity. Our future work will focus on extensions of the DPD system to multi-user hybrid MIMO and multi-user digital MIMO transmitters.}

\bibliographystyle{IEEEbib}
\bibliography{TMTT-2019-12-1447_Ref}

\end{document}